\documentclass[12pt]{article}
\usepackage{graphicx}
\usepackage{multicol}
\usepackage{natbib}

\begin{document}
\pagestyle{myheadings}
\bibliographystyle{apalike}

\linespread{1}

%
%

\begin{center} \textbf{\large Stimulus-invariant processing and
spectrotemporal reverse correlation in primary auditory
cortex}\\[.2in]

David~J.~Klein\textsuperscript{*\textdagger\textdollar}
Jonathan~Z.~Simon\textsuperscript{\textdagger\textsection}
Didier~A.~Depireux\textsuperscript{*\textdaggerdbl}\ and
Shihab~A.~Shamma\textsuperscript{*\textdagger}\\[.2in]

\textsuperscript{*}Institute for Systems Research\\
\textsuperscript{\textdagger}Department of Electrical and Computer Engineering\\
\textsuperscript{\textsection}Department of Biology\\
University of Maryland\\
College Park MD {20742}, USA\\
\vspace{.25in}
\textsuperscript{\textdaggerdbl}Department of Anatomy and Neurobiology\\
University of Maryland\\
Baltimore MD {21201}, USA\\
\vspace{.25in}
\textsuperscript{\textdollar}Institute for Neuroinformatics\\
University/ETH Z\"{u}rich\\
{8057} Z\"{u}rich, Switzerland\\

\end{center}

\vspace{1in}

%
%

\pagebreak
\linespread{1.0}

\begin{spacing}{1.00}

\begin{abstract}

The spectrotemporal receptive field (STRF) provides a versatile and
integrated, spectral and temporal, functional characterization of single
cells in primary auditory cortex (AI). In this paper, we explore the origin
of, and relationship between, different ways of measuring and analyzing an
STRF. We demonstrate that STRFs measured using a spectrotemporally diverse
array of broadband stimuli --- such as dynamic ripples, spectrotemporally
white noise, and temporally orthogonal ripple combinations (TORCs) --- are
very similar, confirming earlier findings that the STRF is a robust linear
descriptor of the cell. We also present a new deterministic analysis
framework that employs the Fourier series to describe the spectrotemporal
modulations contained in the stimuli and responses. Additional insights
into the STRF measurements, including the nature and interpretation of
measurement errors, is presented using the Fourier transform, coupled to
singular-value decomposition (SVD), and variability analyses including
bootstrap. The results promote the utility of the STRF as a core functional
descriptor of neurons in AI.

\end{abstract}


\pagebreak

%
%

\section{Introduction}
\label{sec:intro}

It has been over twenty years since the {\em spectrotemporal receptive
field} (STRF) was conceived to describe and measure auditory neurons' joint
sensitivity to the spectral and temporal dimensions of acoustical energy
\citep{hermes, strf, smolders, eggsep, johannesma}. It was specifically
associated with (1) stimuli characterized by randomly varying
spectrotemporal features, and (2) an approach labeled {\em reverse
correlation}, by which the neuron informs the experimenter, via action
potentials, of the features that were of interest to it \citep{deboerrev,
eggrev}. The STRF offered a view of neuronal function that complemented,
and was usually consistent with, that obtained using classical stimuli such
as tones (tuning curves and rate-level functions), clicks (impulse
responses), and noise (bandwidth sensitivity). In addition, it neatly fit
within an analytical framework, bolstered by the fields of time-frequency
analysis \citep{cohen} and nonlinear systems theory \citep{eggermont},
within which the functionality of neurons could, in principle, be
systematically explored to any level of detail.

The term ``STRF'' does {\it not} denote here the full complex (likely
nonlinear) receptive field of an auditory neuron.  Rather it is a
technical term that has traditionally been used to refer specifically
to the {\it linear} relationship between the time-dependent spike rate
of a neuron and the time- and frequency-dependent energy --- in short,
the {\em dynamic spectrum} --- of a stimulus. In order to measure the
STRF, the reverse-correlation approach prescribes computing the average
dynamic spectrum of those portions of a stimulus preceding the neuron's
spikes. In this context, the STRF is commonly interpreted as the
spectrotemporal pattern that optimally activates a neuron
\citep{youngsci}. Theoretically, as long as all patterns occur
randomly, independently, and equiprobably, the STRF can be revealed by
this ``spike-triggered average'' \citep{eggermont}.

As with tuning curves, rate-level functions, and other commonly used
neuronal response measures, the STRF provides only a limited view of the
receptive field of a neuron, one that is useful only within the context
of the experiment or the nature of information sought from it. For
example, tuning curves are useful as approximate indicators of a unitÕs
BF and bandwidth, but are largely irrelevant as a gauge of its dynamic
range and temporal properties. Similarly, the STRF is a useful measure
of spectrotemporal features likely to drive a cellÕs responses. However,
being a measure of the linear component of the stimulus-response
relationship, it is mostly effective in predicting the linear aspects of
the responses, predictions that can be accurate if the non-linear
portions are small or are well known and can be accounted for in the
measurement (e.g., spike-rate rectification and saturation). In some
cases, the linear component of the response is small and hence one does
not expect clean and reliable STRF measurements, i.e., the STRFs exhibit
significant randomness or high variability across presentations, or are
poor predictors of responses to novel stimuli. Examining these sources
of variability and prediction errors provides useful information
regarding the limitations of the STRF and ways to extend it beyond the
linear domain.

Although the STRF has been slow to mature, it is now increasingly used to
study the physiology of central auditory neurons. In retrospect, the often
slow pace of progress can be partially attributed to the
reverse-correlation methodology, which remains fairly opaque. In
particular, reverse correlation provides no straightforward formal basis
for describing the effectiveness of, or relations between, specific
stimuli, because only the average statistics of stimuli are specified. For
example, Gaussian broad-band noise, the ``ideal'' stimulus for
reverse-correlation, is often ineffective when applied to central auditory
neurons (but see \citep{takahashi}).  Meanwhile, a range of other stimuli
and associated techniques have been auditioned, modulated broad-band noise
\citep{miller1, escabi1}, random sequences of tones or chords
\citep{aertcomp, epping, schafer, decharms, theunissen, rutkowski}, and
natural stimuli \citep{aertcomp, yesh2, schafer, theunissen, sen1}.  While
it is sometimes implied that the auditory system processes different
stimuli differently, it has not been made clear, because of the lack of
vocabulary, to what extent different stimulation methods {\em should} yield
different results.  Additionally, most of the employed stimuli share
randomness in their spectrotemporal design, in accordance with the
reverse-correlation approach, but this style of stimulation is bound to be
inefficient \citep{victknight, sutter}.

Because of these shortcomings, we endeavored to record a {\em deterministic
and analytical} reformulation of spectrotemporal reverse correlation
\citep{robust}. The roots of this new methodology are in the Fourier-based
analysis \citep{papoulis} of any {\em given} stimulus in terms of its {\em
spectrotemporal modulation frequency} content.  Each spectrotemporal
modulation frequency is the conjunction of a spectral and a temporal
modulation frequency; the higher the spectral modulation frequency, the
sharper the spectral feature (e.g., sharp peaks or edges in the spectrum),
and the higher the temporal modulation frequency, the more abruptly that
feature changes in time.  As a population, the strongest phase-locked
response in central auditory neurons occurs over a select range of low
spectral and temporal modulation frequencies \citep{rees1, versnel,
schreirip, kowalski1, dep1, sen1, miller1, escabi1}. Not surprisingly, the
most fruitful stimuli have had spectrotemporal modulation frequencies
concentrated within this range. Our approach extends these past successes
by making explicit the relations between the spectrotemporal modulation
frequency content of a stimulus, the stimulus duration and bandwidth, and
the accuracy of the STRF measurement. This enables the flexible design of
diverse stimuli that minimize both stimulation time and measurement error,
within the constraints of a particular experiment. These constraints
include information about not only the STRF, but also about the nonlinear
and stochastic aspects of the stimulus-response transformation, which are
not directly described by the STRF. Another important advantage of this
methodology is that it can be used to describe the mechanics of STRF
measurement {\em with any given stimulus}, thus providing a language with
which apparently disparate methods can be discussed.

We focus in this article on three specific types of stimuli with increasing
level of complexity, applied in primary auditory cortex (AI) of the
anesthetized ferret. At one extreme are the {\em dynamic ripple stimuli}
\citep{kowalski1, kowalski2, dep1}, which each consist of a single
spectrotemporal modulation frequency. At the other extreme is {\em
spectrotemporally white noise} (STWN), which contains many superimposed
spectrotemporal modulation frequencies. Intermediate are {\em temporally
orthogonal ripple combinations} (TORCs), consisting of special combinations
of several spectrotemporal modulation frequencies each. We shall explore
the relations between these stimuli, and compare the responses they evoke
and the resulting STRF measurements. Among the issues addressed are the
similarity between the STRF measurements, their fidelity and
noise-robustness, their susceptibility to common neuronal nonlinearities,
and the expected amount of data necessary to achieve an measurement with a
desired level of accuracy.  The methods used to address these issues are
quite general, though the specific findings apply only to the population of
neurons in AI studied here.

\section{Methods}

\subsection{Theory}
\label{sec:methodol}

In this section, we outline the methodological basis of this study.  Its
key element is an analytical description of the stimulus-to-response
transformation, in terms of the processing of spectrotemporal modulation
frequencies. In this context, the result of reverse correlation is derived,
first assuming that the response is deterministically and linearly related
to the stimulus, and then considering the separate effects of response
variability and nonlinearity.

At the core of the STRF-based model of neural functionality is the
following equation:
\begin{equation} r(t) = \int{\int {h(\tau,x)
\cdot s(t-\tau,x)\, d\tau}\, dx},
\label{eq:strf}
\end{equation}

where the neuronal response $r$ at any time $t$ is the linear integration
of influences arising from stimulus energy $s$ at different tonotopic
locations $x$ (here corresponding to the logarithm of frequency) and
different times in the past $\tau$. The strength and nature of the
influences --- whether they are excitatory (positive), or suppressive or
inhibitory (negative) --- is described by the STRF as denoted by
$h(\tau,x)$.  In the context of reverse correlation, $r(t)$ is typically
taken to be the time-dependent spike rate of a neuron \citep{eggpred,
takahashi, sen1}.

\subsubsection{The Linear Processing of Spectrotemporal Modulation
Frequencies}
\label{sec:fs}

Our analytical description of dynamic spectra is based upon the {\em
Fourier series} \citep{papoulis}, using elemental Fourier components which are cosine waves as a
function of both $t$ and $x$: $a\cdot cos(2\pi wt + 2\pi\Omega x +
\psi)$. The wave has a peak value of $a$ and starting phase $\psi$. The wave frequency is $w$ cycles/second (Hz) along $t$ and
$\Omega$ cycles/octave (cyc/oct) along $x$. Since the dynamic spectrum
details the modulation of acoustic energy as a function of both $x$
and $t$, these frequencies are referred to as {\em modulation
frequencies}: spectral ($\Omega$) and temporal ($w$). A single Fourier
component is said to consist of a single {\em spectrotemporal
modulation frequency}, defined by a specific $(w,\Omega)$ pair. Just
as a sum of pure tones of various frequencies, amplitudes, and phases
can describe any acoustic waveform over a finite duration, a sum of
various spectrotemporal modulation frequencies (with appropriate
amplitudes and phases) can describe any dynamic spectrum over a finite
duration $T$ and bandwidth $X$. Further, just as the frequency content
of an acoustic waveform (i.e., the amplitudes and phases of its
constituent tones) is described by its (Fourier) spectrum, the
spectrotemporal modulation frequency content of a dynamic spectrum is
described by its {\em spectrotemporal modulation spectrum} $S$.

When the STRF is recast as operating upon $S$, one arrives at a
complementary description called the {\em spectrotemporal modulation
transfer function} $H$. $H[w,\Omega]$, which is the 2-D Fourier transform
of the STRF $h(t,x)$, details the linear component of neural processing of
spectrotemporal modulation frequencies. Such processing is already under
study in auditory neurophysiology \citep{kowalski1, kowalski2, dep1,
miller2, miller1, escabi1} and psychoacoustics \citep{chi}, and is also
being investigated for various signal-processing tasks, including audio
coding \citep{atlas1, sparse} and speech recognition \citep{hermansky1,
nadeu, kleinschmidt1, kleinschmidt2}.

$S$ and $H$ are mathematically
defined as follows. Consider a dynamic spectrum $s(t,x)$ 
and an STRF $h(t,x)$, both
given over a finite range of $T$ seconds and $X$ octaves. Using the
exponential form of the Fourier series, $s$ can be expressed by the
sum
\begin{equation}
s(t,x) = \sum\limits_{k=-\infty}^{\infty}
\sum\limits_{l=-\infty}^{\infty}
\left(a[w_k,\Omega_l]e^{j\psi[w_k,\Omega_l]}\right)
e^{j2\pi(w_kt+\Omega_lx)},
\label{eq:fs}
\end{equation}
where $e$ is the base of the natural logarithm, $j=\sqrt{-1}$, $k$ and $l$
are integers, $w_k=k/T$, and $\Omega_l=l/X$. This is perhaps the simplest
form of the Fourier series to use; ironically it employs ``complex''
exponential functions. These functions are related to the real-valued
Fourier components through the trigonometric identity 
$cos(\phi)= \frac{1}{2} 
\left(e^{j\phi}+ e^{-j\phi}\right)$, etc.
Accordingly, each term in this sum, indexed by $k$ and $l$, has a 
complex-conjugate counterpart, indexed by $-k$ and $-l$, such that 
$a[w_k,\Omega_l]= a[w_{-k},\Omega_{-l}]$ and $\psi[w_k,\Omega_l]= 
-\psi[w_{-k},\Omega_{-l}]$. Henceforth we will simplify the notation by 
dropping the $k$ and $l$ subscripts, however keeping in mind that $w$ and 
$\Omega$ are discrete-valued variables (as indicated by the square brackets). 
Thus, the amplitudes and phases of the modulation-frequency components are 
given by $a[w,\Omega]$ and $\psi[w,\Omega]$, which together form 
$S[w,\Omega]= a[w,\Omega]e^{j\psi[w,\Omega]}$. As for the STRF, its Fourier 
series description can be represented by the same sinusoidal components, but 
with different amplitudes $b[w,\Omega]$ and phases $\theta[w,\Omega]$, which 
together form $H[w,\Omega]= b[w,\Omega] e^{j\theta[w,\Omega]}$.  As we'll 
see,  $b$ generally describes describes the strength of the response to 
particular spectrotemporal modulation frequencies, while $\theta$ describes 
the timing of the response.

In practice, $s(t,x)$ is represented on a computer by discrete
samples, $s[t_k,x_l]=s(k\Delta_t,l\Delta_x)$, taken at a rate of
$1/\Delta_t$ samples/second and $1/\Delta_x$ samples/octave, where $k$
and $l$ are integers. Again, we will drop the $k$ and $l$ subscripts,
however keeping in mind that $t$ and $x$ are now discrete-valued
variables. By the sampling theorem \citep{oppenheim}, this assumes
that $S$ is sufficiently smooth; that is, it can be described by a
limited number of temporal and spectral modulation frequencies no
higher than $1/(2\Delta_t)$ and $1/(2\Delta_x)$, respectively. Within
these limits, $S[w,\Omega]$ is then obtained by computing the
Discrete Fourier Transform (DFT) of $s[t,x]$ (using the Fast Fourier
Transform, or FFT, algorithm) \citep{oppenheim}. Analogously,
$H[w,\Omega]$ is obtainable via the (Discrete) Fourier
Transform of the STRF $h[t,x]$. 

Since the response, $r(t)$, depends only on time, its Fourier-series
description utilizes only temporal modulation frequencies. It can be
derived by inserting the Fourier-series descriptions of $s$ and $h$
into Eq. (\ref{eq:strf}) and carrying out the integration. The
result is that the Fourier Transform of the sampled
response $r[t]$ has the form
\begin{equation}
R[w] = \sum\limits_\Omega{ H[w,-\Omega]\cdot
S[w,\Omega] } = \sum\limits_\Omega{ H[w,\Omega]\cdot
S[w,-\Omega] }
\label{eq:re}
\end{equation}

Recall that in Eq. (\ref{eq:strf}) the response was obtained by integrating 
over the spectral axis ($x$) after temporally convolving the dynamic spectrum 
with the STRF; here, the convolution is realized via the multiplication of 
Fourier Transforms\footnote{Strictly speaking, this implements a circular 
convolution. If the stimulus is not periodic, this can be converted to a 
linear convolution by including zeros (silence) before and after the stimulus 
\citep{oppenheim}.} \citep{oppenheim}, and the integration over $x$ is 
replaced by a summation over $\Omega$ . Therefore, each frequency $w$ in the 
response results from all spectrotemporal modulation frequencies in the 
stimulus sharing the same temporal component $w$.

\subsubsection{Fourier-based Reformulation of Spectrotemporal Reverse
Correlation}
\label{sec:rc}

The STRF was, in Section \ref{sec:fs}, recast in terms of the processing
of spectrotemporal modulation frequencies. The result of
spectrotemporal reverse correlation will now be derived in this
context.

If spike times are quantized, and stimuli are sampled, with a temporal
resolution $\Delta_t$, then the average stimulus preceding a neuron's
spikes is proportional to the temporal cross-correlation between the
stimulus and a ``binned spike train'' response, $y[t]$, consisting of the
number of spikes observed in consecutive $\Delta_t$ intervals
\citep{eggrev}. For now, we assume that $y[t]/\Delta_t$, with units of
spike rate (spikes/second), is equal to $r[t]$ (the sampled STRF-based
response), whose Fourier Transform $R[w]$ was derived in Eq.
(\ref{eq:re}). Cross-correlation is a linear operation and, much like
convolution, it can be realized via the multiplication of Fourier
Transforms\footnote{Modulo the previous note concerning circular
convolution} \citep{oppenheim}. This takes the following form, in the case
of spectrotemporal reverse correlation:
\begin{eqnarray}
R[w]\cdot S^\ast[w,-\Omega]
& = & H[w,\Omega]\cdot |S[w,-\Omega]|^2 \nonumber \\
& & + \sum\limits_{\Omega^\prime\ne\Omega} {H[w,\Omega^\prime]
S[w,-\Omega^\prime] S^\ast[w,-\Omega] } \nonumber \\
& = & H[w,\Omega]{\cdot} (a[w,-\Omega])^2 +
\tilde{\epsilon}[w,\Omega],
\label{eq:rc}
\end{eqnarray}
where ${}^\ast$ denotes complex conjugation and $|S[w,\Omega]|=
\sqrt{S[w,\Omega]\cdot S^\ast[w,\Omega]}= a[w,\Omega]$ is
the magnitude of $S$. Eq. (\ref{eq:rc}) represents the
Fourier Transform of the reverse correlation result.

An important special case exists when $|S|$ is flat ($a[w,\Omega]=a$) over
the extent of $H$ that is nonzero, and further
$\tilde{\epsilon}[w,\Omega]=0$. Then, Eq.  (\ref{eq:rc}) is proportional to
the $H$, with
\begin{equation}
H[w,\Omega] = {\frac{R[w]\cdot S^\ast[w,-\Omega]}{a^2}},
\label{eq:tfest}
\end{equation}
Since $h[t,x]$ is, by definition, the inverse Fourier Transform of
$H[w,\Omega]$, this implies that, in this special case, reverse correlation
will yield a result proportional to the STRF.

This desirable result has immediate implications for effective stimulus
design. That the spectrotemporal modulation spectrum should be flat
equivalently requires the stimulus contain in equal strength all
spectrotemporal modulation frequencies needed to construct $H$. If the
stimulus contains a subset of the necessary modulation frequencies, then
only part of $H$ can be constructed:  $H$ will be filtered. The
$\tilde{\epsilon}=0$ requirement is not so simply related.  This is a {\em
systematic stimulus-induced error}, dependent upon temporal correlations
between different spectrotemporal modulation frequencies in the stimulus
(it may also be framed in terms of temporal correlations between the
stimulus energy at different tonotopic locations) \citep{robust,
theunissen}. It will be nonzero if the stimulus contains multiple
spectrotemporal modulation frequencies that share the same value of $|w|$,
and therefore by Eq. (\ref{eq:re}) evoke the same frequency in the
response. For a general stimulus, $\tilde{\epsilon}$ will {\em not} be
zero, or even small, and therefore one of three methods must be used to
eliminate or reduce its effects: First, if stimuli are sufficiently diverse
over time or over multiple stimuli, then $\tilde{\epsilon}$ asymptotically
approaches zero as the stimulus duration or the number of stimuli increases
\citep{robust}; second, specially designed stimuli may be employed for
which $\tilde{\epsilon}$ is zero \citep{kvale, robust}; and third,
additional computations may be undertaken to try and adjust for the
correlations in the stimulus \citep{aertsen2, aertcomp, theunissen}. In
this article, we concentrate on the first two of these methods.

Given some knowledge about $H$, creative stimulus design is facilitated by
the simple relationship of Eq. (\ref{eq:tfest}) between the measurement of
points in $H$ and the corresponding points in the spectrotemporal
modulation spectrum.  For example, suppose $H$ is {\em quadrant-separable}
\citep{kowalski2, dep1}, i.e., within each quadrant, the value at every
point is the product of a single vertical cross-section with a single
horizontal cross-section. Then, using only stimuli from a single vertical
cross-section and a single horizontal cross-section within each quadrant is
sufficient to measure the entire $H$. As discussed below, the assumption of
quadrant separability is made for STRFs measured using one stimulus set
(dynamic ripples). Note that the same measurements could be made using
differently structured stimuli that directly probe all points of $H$. The
extent that measured STRFs agree across stimulus sets measures linearity;
but the extent that STRFs measured using dynamic ripples disagree with the
other measured STRFs, does not distinguish between lack of linearity and
lack of quadrant separability.

Thus far, we have assumed that the response is deterministically and
linearly related to the dynamic spectrum. In the next two sections, we
relax these assumptions and consider how response variability and
nonlinearity effects the real-world results. Accordingly, Eq.
(\ref{eq:tfest}) is henceforth treated as a {\em measurement} of
$H$ (and subsequently the STRF), using an observed
response that is not necessarily fully described by the STRF.

\subsubsection{Reliability of the STRF Measurement}
\label{sec:var}

We have assumed thus far that the transformation from stimulus to
response is deterministic. However, in response to identical stimulus
presentations, neuronal responses exhibit inherent variability
\citep{shadlen}, and so the result of reverse correlation is
somewhat indeterminate. Therefore, Eq. (\ref{eq:rc}) should be
interpreted as the {\em mean} result, which would be obtained by
averaging the results of an infinite number of identical experiments.
Due to the linearity of reverse correlation, this is also the result
obtained if $r[t]$ is taken to be the {\em mean} of $y[t]/\Delta_t$
(the mean time-dependent spike rate).

This mean result is called the {\em signal}. The difference between
the actual measurement and its mean is called {\em noise}. The exact
form of the noise varies from measurement to measurement. The mean
squared-magnitude of the noise, as a function of $t$ and $x$, is
called the {\em variance} of the measurement (the square of the
standard error). The overall reliability of the measurement can be
gauged from the {\em signal-to-noise ratio},
$SNR=P/\left<\sigma^2\right>$, which is the average power
(squared-magnitude) of the signal ($P$) relative to the average
variance of the noise ($\left<\sigma^2\right>$), where the averages
are performed over all $t$ and $x$. Note that both $P$ and
$\left<\sigma^2\right>$ are preserved by the Fourier Transform
\citep{papoulis, oppenheim}, and therefore the $SNR$ of $h[t,x]$ is
identical to that of $H[w,\Omega]$ (with the averages performed
over $w$ and $\Omega$).

With this in mind, the signal and noise components of the SNR can be
directly traced through Eq. (\ref{eq:tfest}) to the response. The
variance of $H$ is found to be
\begin{equation}
Var\left\{H[w,\Omega]\right\} =
{\frac{Var\left\{R[w]\right\}\left|S[w,-\Omega]\right|^2}{a^4}}
= {\frac{Var\left\{R[w]\right\}}{a^2}},
\label{eq:tfvar}
\end{equation}
since $R[w]$ is the only source of variance.

Analogously, the squared-magnitude (power) of $H$ is
\begin{equation}
\left|H[w,\Omega]\right|^2 =
{\frac{\left|R[w]\right|^2}{a^2}}.
\end{equation}
If $r$ is taken to be the {\em mean} response, this equation describes
the {\em signal} power. If instead $r$ denotes the {\em actual}
response, then the resulting $H$ {\em measurement} (and
equivalently, the STRF measurement) will be composed of signal plus
noise, and therefore its average power will exceed $P$ by
$\left<\sigma^2\right>$, provided the signal and noise components are
uncorrelated.

In summary, response variability is a source of error in the STRF
measurement. This is referred to as {\em non-systematic error}, since
its exact form varies from measurement to measurement. The expected
size of the error is quantified by $\left<\sigma^2\right>$. At the
same time, the signal power ($P$) and response power are closely
related. Therefore, stimuli that maximize the response power relative
to the response variance will result in more reliable STRF measurements
(higher SNR). Note also that, in theory, the SNR of the STRF
measurement could be obtained directly from the response, without
actually computing the STRF.

\subsubsection{Nonlinear Contributions}

So far, we have only discussed the relationship between modulations in the
dynamic spectrum and modulations of the mean spike rate as being purely
linear. Of course nonlinearities such as rectification (the strictly
positive nature of the spike rate) and synaptic depression \citep{chance1,
carandini1} introduce additional response components.  To the extent that
these components are correlated with the stimulus, they result in {\em
systematic}, {\em stimulus-dependent errors} to the STRF measurement.

A detailed accounting for various nonlinearities is not given here.
Suffice it to say that a portion of the response can be described by
Eq. (\ref{eq:strf}), and the remaining nonlinear portion may be
described by additional terms in a Volterra or Wiener functional
expansion, which have long been used in neuroscience \citep{eggermont}
and systems theory \citep{schetzen}. The portion of the nonlinearity
manifest at the odd- and even-numbered terms of the expansions is
dubbed odd- and even-order nonlinearity, respectively. Fourier-based
descriptions of the input-output characteristics of such systems are
already well studied (e.g., \citep{victknight, victor, boyd}). They
describe how multiple stimulus frequencies (e.g., spectrotemporal
modulation frequencies) interact to form nonlinear response
frequencies, or {\em distortion products}. It is those distortion
products manifested at frequencies overlapping with the linear portion
of the response that interfere with the STRF measurement.

Knowledge about the stimulus dependence of distortion products
facilitates the detection, identification, and extraction of nonlinear
response elements \citep{spekreijse, victor, boyd}. For example, odd-
and even-order nonlinearities are distinct in that their distortion
products are composed of products of odd and even numbers of stimulus
elements, respectively. By straightforward trigonometry, one can
determine the possible response frequencies that may be observed for a
stimulus of known (or cleverly designed) composition, and further
determine how the amplitude of these distortion products will change
if a gain is applied to the stimulus.

\subsection{Experimental Details}

We now detail how the above methodology is exploited by the methods
used in this study.

\subsubsection{Surgery and animal preparation}

Data were collected from 16 domestic ferrets (\textit{Mustela
putorius}) supplied by Marshall Farms (Rochester, NY). The ferrets
were anesthetized with sodium pentobarbital (40 mg/kg) and maintained
under deep anesthesia during the surgery. Once the recording session
started, a combination of Ketamine (8 mg/Kg/Hr), Xylazine (1.6
mg/Kg/Hr), Atropine (10 $\mu$g/Kg/Hr) and Dexamethasone (40
$\mu$g/Kg/Hr) was given throughout the experiment by continuous
intravenous infusion, together with Dextrose, 5\% in Ringer solution,
at a rate of 1 cc/Kg/Hr, to maintain metabolic stability. The
ectosylvian gyrus, which includes the primary auditory cortex, was
exposed by craniotomy and the dura was reflected. The contralateral
ear canal was exposed and partly resected, and a cone-shaped speculum
containing a miniature speaker (Sony MDR-E464) was sutured to the
meatal stump. For more details on the surgery see \citep{sasorg}.

\subsubsection{Recordings, spike sorting, and selection criteria}

Action potentials from single units were recorded using
glass-insulated tungsten microelectrodes with $5$--$7$ M$\Omega$ tip
impedance at $1$ kHz. In each animal, electrode penetrations were
made orthogonal to the cortical surface. In each penetration, cells
were typically isolated at depths of $350$--$600$ $\mu$m corresponding
to cortical layers III and IV \citep{sasorg}. In $12$ animals, neural
signals were fed through a window discriminator and the time of spike
occurrence relative to stimulus delivery was stored using a computer.
In the other $4$ animals, the neural signals were stored for further
processing offline. Using MATLAB software designed in-house, action
potentials were then manually classified as belonging to one or more
distinct neurons, and the spike times for each neuron were recorded.
The action potentials assigned to a single neuron met the following
criteria: (1) the peaks of the spike waveforms exceeded $4$ times the
standard deviation of the entire recording; (2) each spike waveform
was less than $2$ ms in duration and consisted of a clear positive
deflection followed immediately by a negative deflection; (3) the
spike waveforms were not visibly different from each other, modulo the
noise; (4) the histogram of inter-spike-intervals evidenced a minimum
time between spikes (refractory period) of at least $1$ ms. This
procedure occasionally produced units with very low spike counts.
After consulting the distribution of spike counts for all units, units
that fired fewer than one spike per two seconds of stimulation were
excluded from further analysis.

Analysis of the dynamic-ripple recordings was published previously
\citep{dep1}. Here we used the same selection criteria for those
recordings that were used in that study. Those criteria were somewhat
more stringent than those used for the TORC and STWN recordings;
consequently, there are conspicuously fewer instances of low-SNR
STRFs and low spike counts in the dynamic-ripple results, with respect
to the TORC and STWN results.

\subsubsection{Stimulus Realization and Delivery}

A stimulus is designed by first specifying its envelope $S$.
Recall from Section \ref{sec:rc} that the spectrotemporal modulation
frequencies contained in the stimulus are used to reconstruct the
STRF. Through the properties of the Fourier Series described in
Section \ref{sec:fs}, the set of frequencies required for this
construction is defined by four parameters: $T$ and $X$, the temporal
extent (memory) and spectral extent (bandwidth) of STRF; and $w_c$ and
$\Omega_c$, the maximum temporal and spectral modulation frequencies
in $H$. For all results reported here, $T$ was $250$ ms,
$X$ was $5$ octaves, $w_c$ was $24$ Hz, and $\Omega_c$ was $1.4$
cyc/oct. These values were chosen {\em a priori} based upon the likely
structure of STRFs in AI, as inferred from previous studies
\citep{kowalski1, kowalski2, dep1}.

The requisite set of modulation frequencies need not be contained
within a single stimulus; it may be divided among multiple stimuli.
Stimuli thus devised are used to independently reconstruct different
areas of $H$, which are finally combined to form the
complete measurement. Some benefits of this scheme include the
reduction of measurement errors and the option of using short-duration
stimuli \citep{robust}.

The design of $S$ subsequently specifies (via an inverse
Fourier Transform) a desired or ``target'' dynamic spectrum. We
realized this target with a sum of amplitude-modulated (AM) tones of
various carrier frequencies (typically $100$ tones per octave) and
random phases \citep{kowalski1}. First, the target is scaled so that
its values lie within $\pm90\%$ of the mean value. The mean value,
which corresponds to the mean amplitude of the tones, is set
$10$--$20$ dB above the neuron's threshold (measured previously with
pure tones). Finally, the AM pattern of each tone is specified by the
cross-section of the envelope $S$ at the corresponding spectral location $x$.

\begin{figure}[!ht]
\begin{center}
\includegraphics[width=.65\linewidth]{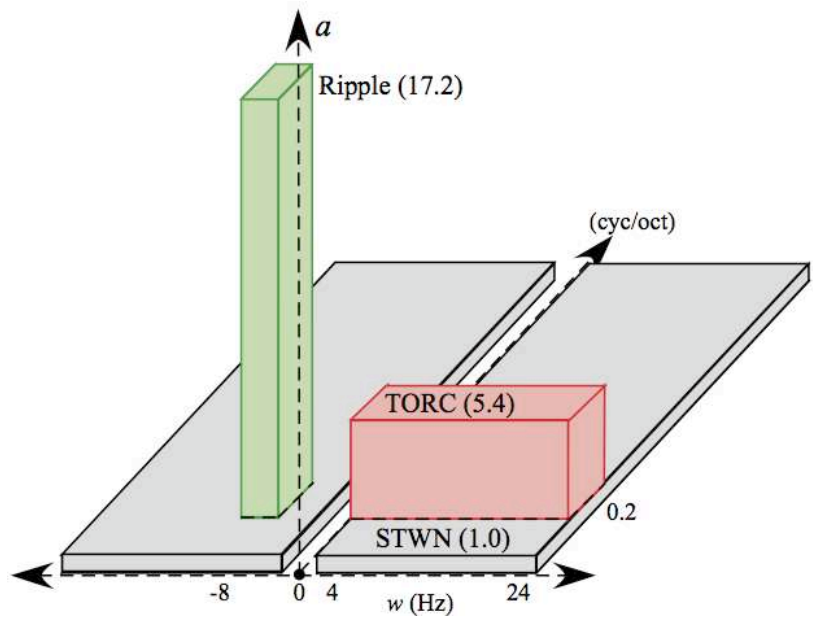}
\end{center}
\caption[Modulation frequency amplitudes]{\footnotesize The
$S$ magnitudes are illustrated for members of each of the
three stimulus types --- dynamic-ripple stimuli, TORCs, and STWN. The
stimuli all have the same duration ($250$ ms), and contain $1$, $6$,
and $90$ spectrotemporal modulation frequencies, respectively. By
virtue of the dynamic range constraint on the intensities of the
dynamic spectrum, the stimuli must employ different
modulation-frequency amplitudes $a$. The amplitudes, relative
to those of the STWN stimulus, are indicated in parentheses.}
\label{fig:as}
\end{figure}

Three types of stimuli are used in this study: dynamic-ripple stimuli,
temporally orthogonal ripple combinations (TORCs), and
spectrotemporally white noise (STWN). As exemplified in Figure
\ref{fig:as}, they distribute spectrotemporal modulation frequencies
among stimuli in different ways. Due to the peak-amplitude constraint
on the dynamic spectra, they also employ markedly different
modulation-frequency amplitudes; increasing the number of
modulation frequencies in a stimulus (implying more complex
modulations) generally requires the amplitude of each frequency to be
decreased so that their sum is contained within a given range. In any
case, the amplitudes of all modulation frequencies within a given
stimulus were identical. If a stimulus contained multiple modulation
frequencies, their phases were randomly assigned; otherwise they were
(arbitrarily) set to zero. Additional details about these stimuli are
provided later in Section \ref{sec:over}.

\end{spacing}

\begin{spacing}{.9}
\end{spacing}
\begin{spacing}{1.00}

The Fourier series endows dynamic spectra, thus designed, with a
common periodicity of $T=250$ ms and $X=5$ octaves. One spectral
period was realized in each stimulus, whose $5$-octave bandwidth was
centered upon the neuron's pure-tone tuning curve (measured
previously). The temporal periodicity of the dynamic spectra was
exploited; this enabled multiple observations of the response, since
(assuming the neuron's memory is less than $T$ seconds) all temporal
periods beyond the first constitute identical stimulus presentations.
A stimulus sweep consisted of a limited number ($4$ or $12$) of
stimulus periods, and had a rise and fall time of $8$ ms. Multiple
sweeps were presented for each stimulus. Sweeps of different stimuli,
separated by $3$--$4$ seconds of silence, were presented in a
pseudorandom order, until a neuron was exposed $60$--$120$ periods
($15$--$30$ s) of each stimulus.

All stimuli were gated and fed through an equalizer into an earphone.
Calibration of the sound delivery system (to obtain a flat frequency
response up to $20$ kHz) was performed in situ with the use of a $1/8$
in.\ Br\"uel \& Kjaer 4170 probe microphone. The earphone was inserted
into the ear canal through the wall of the speculum to within $5$ mm
of the tympanic membrane. The speculum and microphone setup resembles
closely that suggested by Evans \citep{evans}. 

\subsubsection{Response Measurement and STRF Calculation}

Each stimulus resulted in a collection of response observations
${y[t]}$ (i.e., binned spike trains), each member of which consisted of
the number of spikes occurring in successive $\Delta_t=1$ ms intervals
during one stimulus period (see, e.g., Figure \ref{fig:dr}B). The
total number of stimulus periods used was $n$. The transient epochs,
during the first period of each sweep, were disregarded; only the
steady-state portion of the response was utilized. The spike rate
$r[t]$ was then estimated from the sample mean of $y[t]/\Delta_t$:
$r[t]={\frac{1}{n}}
\sum\limits_{i=1}^n{y_i[t]/\Delta_t}$, where
$y_i[t]$ is the response to the $i$\textsuperscript{th} stimulus
period. This is the response whose Fourier Transform is used to
calculate $H$ (and subsequently the STRF), or some
portion thereof, via Eq. (\ref{eq:tfest}). These calculations are very
simple and are completed in MATLAB (Mathworks) in a fraction of a
second.

\subsubsection{Reducing Nonlinear Interference with the Inverse-Repeat
Method}

In this article, we concentrate on even-order nonlinearities; they
are ubiquitous in the brain (e.g., due to rectification), and can
severely distort the reverse-correlation measurement, particularly
when the stimulus is brief \citep{swerup}. Fortunately, its ill effects
are easily isolated and extracted by the inverse-repeat
method \citep{moller2, wickesberg}. In its simplest form, this method
calls for two stimuli (here, dynamic spectra) that sum to a constant
value. While the linear responses to the two stimuli are opposite in
sign, the even-ordered distortion products are identical
\citep{victor}. Therefore, the even-order effects are removed by
subtracting the two responses and dividing by two (or instead isolated
by adding the responses). This method is investigated in conjunction
with TORC stimulation.

\subsubsection{Signal and Noise Calculations: Non-systematic Errors}

As mentioned in Section \ref{sec:var}, the measures of signal power
$P$ and noise variance $\left<\sigma^2\right>$, and therefore
the SNR, apply to both $h[t,x]$ and $H[w,\Omega]$. For a
single stimulus-response pair, a simple relationship was identified in
Eq. (\ref{eq:tfvar}) between the variance of $H[w,\Omega]$ and
the variance of $R[w]$. Note that latter variance is, in turn,
proportional to the variance of $\tilde{y}[w]$, the Fourier Transform
of the response to one stimulus period; specifically,
\begin{equation}
Var\left\{R[w]\right\} ={\frac{1}{n}}
{\frac{Var\left\{\tilde{y}[w] \right\}}{\Delta_t^2}}.
\label{eq:rvar}
\end{equation}
Thus, the variance of $H[w,\Omega]$ could be quickly estimated
from the sample variance of $\tilde{y}[w]$ (across all stimulus
periods), without repeating the experiment or subdividing the data.

However, the $H$ measurement may incorporate the
measurements from multiple stimulus-response pairs; if so, its
variance will depend on how the individual measurements are combined.
If a point on $H[w,\Omega]$ is the average of $N$ measurements,
then its variance will simply tend to scale by $1/N$ with respect to
that of an individual measurement. But more complicated functions of
the individual measurements (such as that used for the dynamic-ripple
stimuli \citep{dep1}) may obscure the relation between the variance of
$H$ and that of the constituent responses. In such a
case, the {\em bootstrap method} may be employed. This method
simulates the randomness of a statistic that is a function of a
collection of identical observations, without repeating the experiment
or subdividing the observations \citep{efron, politis}. In the present
context, a new $H$ is computed from a new,
identical-sized collection of ${y[t]}$, assembled by selecting members
of the original collection randomly and with replacement. The sample
variance of $H$, or some function thereof, is
calculated after repeating the process many times (we used $300$),
which is feasible due to the simplicity of the computations.

For the sake of equal footing, we used the bootstrap method to
estimate the variance of $H$ for all stimulus
types. After subsequently calculating $\left<\sigma^2\right>$, the
$SNR$ was inferred from the average power of 
$H$, which, as mentioned in Section \ref{sec:var},
approximately equals $P+\left<\sigma^2\right>$.

\subsubsection{Signal and Noise Calculations: Systematic Measurement Errors}

The SNR quantifies the size of the signal compared to the size of the
{\em non-systematic} component of the measurement error. However, the
possible additional contribution of {\em systematic} errors --- that
is, those induced by non-ideal stimulus structure (i.e., $\tilde{\epsilon}$
in Eq. \ref{eq:rc}) and by nonlinearities --- cause the actual error
level of the STRF measurement to exceed that described by the SNR. There exists an
opportunity to obtain a more ``correct''
measure of the SNR, provided that all errors are evenly distributed
over the STRF measurement, because the signal tends to be concentrated in
an early region of the STRF measurement between $0$ and $125$ ms -- in other words, neuron's responses are only weakly effected by stimulus conditions more than$125$ ms in the past.
Accordingly, a corrected SNR measure, $SNR_{cor}$, was obtained after
dividing the average power of the early region of the STRF measurement by the average power of
the late (post $125$ ms) region. Note that the late region of the STRF measurement contains the uncorrelated contributions of both non-systematic and systematic errors, while the noise power estimate used for $SNR$ only measures the non-systematic component; therefore, $SNR_{cor}$ should be less
than or equal to $SNR$ (modulo the inaccuracies in measuring $SNR$ and
$SNR_{cor}$), with equality when there are no systematic errors.

\subsubsection{Error Reduction with the Singular-Value Decomposition}

To further reduce errors in the STRF measurement, we investigated the
singular-value decomposition (SVD), applied to either $h[t,x]$ or
$H[w,\Omega]$ (which are both just matrices of numbers). The
SVD is a well-studied tool for resolving the structure of matrices
that are corrupted by errors \citep{stewart1, hansen}. It works by
breaking up an arbitrary matrix into a sum of separable matrices,
which, in the current context, are each formed by the product of one
temporal vector and one spectral vector. The first matrix takes the
best separable approximation out of the original matrix; the second
takes the best separable approximation out of the remainder, and so
on. The importance of each separable matrix is gauged by its singular
value, which is the square root of its average power. The total number
of separable matrices required to describe a matrix (the number of
nonzero singular values) is called the matrix's {\em rank}.

A basic theorem \citep{stewart2} implies that if the error-free STRF
can be well approximated by only a few separable matrices, then the
addition of {\em small and evenly distributed} errors will only
slightly perturb their form, as they constitute the first few matrices
in the SVD of the STRF measurement. The additional and subsequent
matrices required to describe the measurement will describe mostly
errors, and thus should be discarded. In fact, there are {\em a
priori} reasons to believe that STRFs are well approximated by
low-rank matrices. Typically, cortical STRFs are localized in a
compact area of the spectrotemporal domain and the
modulation-frequency domain \citep{dep1, miller1}; this alone will
limit their rank. Still lower limits will be imposed by special
structure within the STRF or the $H$, such as
spectral-temporal separability \citep{eggsep, dep1, sen1}, quadrant
separability \citep{dep1}, and temporal symmetry \citep{tempsep}.

In practice, determining which separable matrices should be discarded
is a complex problem \citep{stewart1, hansen}. Most approaches use
knowledge or assumptions about the size and structure of the errors to
bound the singular values (or functions thereof) of those separable
matrices describing mostly errors. Through simulations, we found that
methods based solely on variability analysis tended to underestimate
the size of the errors; instead, the most generally accurate methods
gauged the error level directly from the post-125-ms region of the
STRF measurement (for a similar method see \citep{sen1}). We used the
largest singular value from this region (or its Fourier Transform) to
threshold the singular values of the pre-125-ms region (or its Fourier
Transform). In theory, the STRF (or $H$) is optimally
approximated using only those separable matrices with singular values
above this threshold, and discarding the remainder \citep{stewart1, hansen}.

Although this approximation is in some sense optimal, it is still
prone to error. As the error level increases, more and more error
leaks into the approximation and, conversely, more and more of the
STRF power is lost under the error threshold \citep{hansen}. This
second case is of primary interest in this study; we will gauge the
proportion of (error-free) STRF power excluded from the SVD
approximation. A naive gauge of this is $\alpha_{SVD}$, the proportion
of the STRF {\em measurement's} power contained in the SVD remainder
\citep{dep1}. Unfortunately, when the level of measurement error is
high, $\alpha_{SVD}$ itself will be inflated, because much of the
remainder will consist of error. However, we can use the bootstrap
method to estimate the size (average variance) of the part of the
remainder resulting from non-systematic errors, and subtract it out.
This leads to a more accurate gauge of the proportion of lost STRF
power, particularly when the systematic errors are small:
$\beta_{SVD}$, the average power of the systematic component of the
remainder, divided by $P$. In Section \ref{sec:svdres}, we use
$\alpha_{SVD}$ and $\beta_{SVD}$ together to study how measurement
errors effect the performance of the SVD.

\subsubsection{STRF Comparisons}

In this article, the correlation coefficient is used to quantify the
similarity between two different STRF measurements. This takes values
between $-1$ and $+1$, with $+1$ indicating a perfect match.  Comparisons
are made over the first $125$ ms of the measurements, both before and after
the SVD is applied. Note that the correlation coefficients for the pre-SVD
comparisons will be limited by $SNR_{cor}$; if two identical STRFs are
corrupted by independent and identically distributed errors, the
correlation coefficient should approximately equal
$SNR_{cor}/(SNR_{cor}+1)$.  To the extent that the SVD approximations
result in increased SNRs, they will allow for higher correlation
coefficients, which we modeled as $gSNR_{cor}/(gSNR_{cor}+1)$, where $g$
represents a multiplicative gain in $SNR_{cor}$.

\subsubsection{Simulations}

Simulations were employed in order to verify the performance of these
methods under realistic conditions. The core of a simulation is an
STRF (tailor-made or derived from a low-rank approximation of an
actual measurement) and a set of stimuli. The STRF-based responses to
the stimuli are computed via Eq. (\ref{eq:re}). These responses are
then altered; usually they are rectified and then subjected to another
static nonlinearity, such as a squaring function. The result,
representing the time-varying spike rate, is used to create spike
trains with inhomogeneous Poisson statistics \citep{berry, oram}, with
a time step of $50$ $\mu$s. These spike trains are treated as the
responses of a neuron with an unknown STRF, and are subjected to the
very same analyses as the real responses. Wherever the bootstrap
method was employed, its expected performance was simulated against a
Monte-Carlo procedure, employing 300 sets of independent responses with
identical spike rates.

\section{Results}
\label{sec:results}

The results of this study are presented as follows. In Section
\ref{sec:over}, we detail the measurement of a neuron's STRF using
each of the three stimulation types, and we subsequently illustrate
the computation of the SVD-based STRF approximations. In Section
\ref{sec:ccs}, for neurons whose STRFs were measured with multiple
stimulus types, we examine the similarity between the multiple
measurements and the corresponding SVD approximations, as a function
of the level of measurement error. In Section \ref{sec:err}, we
analyze the origins and stimulus dependence of the measurement errors.
Finally, in Section \ref{sec:svdres}, we study how measurement errors
affect the sufficiency of the SVD approximations.

\subsection{Overview}
\label{sec:over}

In this section, we detail the measurement of a neuron's STRF using
dynamic-ripple stimuli (Figure \ref{fig:dr}A), TORCs (Figure
\ref{fig:dr}B), and STWN (Figure \ref{fig:dr}C), respectively. The
$S$ magnitudes for examples of each of these stimulus
types are illustrated in Figure \ref{fig:as}. The respective STRF
measurements are denoted STRF$_{{DR}}$,
STRF$_{{TORC}}$, and STRF$_{{STWN}}$. Computation of
the SVD-based approximations of the measurements is subsequently
detailed.

\subsubsection{Dynamic-Ripple Stimuli}
\label{sec:dr}

\begin{figure}[!ht]
\begin{center}
\includegraphics[angle=0,width=\linewidth]{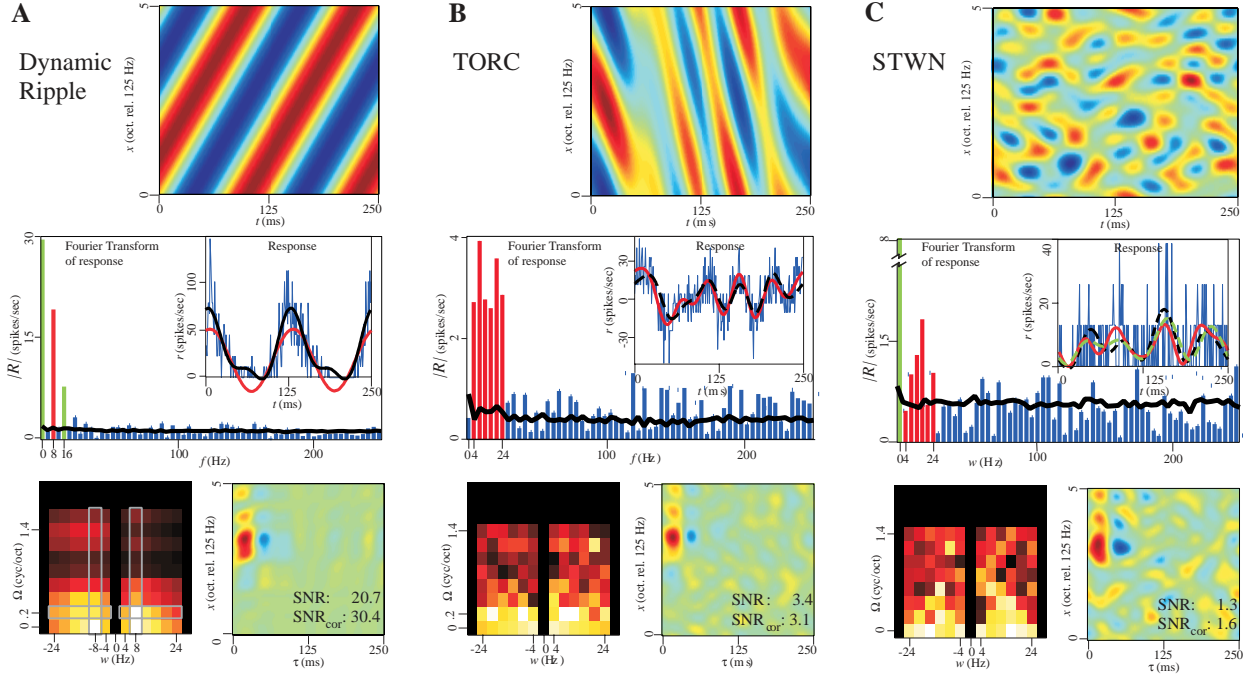}
\end{center}
\caption[STRF Measurements]
{\footnotesize Measuring the STRF 
of one neuron with different types of ripple stimuli. 
A: (Top panel) Dynamic spectrum of a single dynamic-ripple stimulus with 
$w=-8$ Hz and
$\Omega=0.2$ cyc/oct. $90$ stimulus periods were used. (Middle panel)
Inset displays the response as time-dependent
spike rate estimate, $r[t]$: Raw estimate (blue) (using $\Delta_t=1$
ms), linear ($8$ Hz) plus DC ($0$ Hz) approximation (red), and the
approximation obtained by including the (even-order) $16$ Hz
distortion product (black). The magnitude of the Fourier Transform 
of response is shown in the panel,
clearly exhibiting the linear $8$ Hz component (red), nonlinear
distortion products (green), and the remaining noise component (blue).
Also shown is the square-root of the response variance (the standard
error) as a function of frequency (black). (Bottom panels): Measurements of
(left) $H$ and (right) $h$ (or STRF$_{{DR}}$) after all $30$
stimuli. The grey outlines in $H$ (left panel) indicate 
the cross-sections that were directly measured. B: (Top panel) Dynamic spectrum of 
a TORC with $\Omega=0.2$ cyc/oct and $w$'s between $4$ and $24$ Hz. $75$
stimulus periods were used. (Middle panel) Inset shows response as time-dependent 
spike rate estimate, $r[t]$, after the inverse-repeat procedure: Raw
estimate (blue), linear plus DC approximation (red) obtained by
discarding frequencies above $24$ Hz, and the response predicted from
the previously obtained STRF$_{{DR}}$ in A (dashed black). (Bottom panels): 
Same as in A above using 15 pairs of TORC stimuli. C: (Top panel) Dynamic spectrum 
of a STWN with $\Omega$'s between $0.2$ and $1.4$ cyc/oct and $w$'s between $4$ 
and $24$ Hz. $75$ stimulus periods were used. (Middle panel) Inset shows response 
as time-dependent spike rate estimate, $r[t]$: Raw estimate
(blue), the linear plus DC approximation (red), and the response
predicted from STRF$_{{DR}}$ (dashed black) and
STRF$_{{TORC}}$ (dashed green) in A and B above. Panel also shows response 
Fourier Transform magnitude. The linear $4$--$24$ Hz components (red) are barely
distinct from the noise (blue). Also shown is the square-root of the
response variance (the standard error) as a function of frequency
(black). (Bottom Panels) The measurements of $H$ and
STRF$_{{STWN}}$ averaged over $30$ stimuli.}
\label{fig:dr}
\end{figure}

For the dynamic-ripple stimuli \citep{kowalski1, dep1} shown in Figure
\ref{fig:dr}A, each stimulus is composed of a single spectrotemporal
modulation frequency (Fourier component). It can therefore be
considered the auditory equivalent to the drifting sinusoidal
luminance gratings used in visual neuroscience \citep{devalois}. Figure
\ref{fig:dr}A shows the dynamic spectrum of one such stimulus (top panel), which
has a temporal modulation rate $w$ of $-8$ Hz and a spectral
modulation rate $\Omega$ of $0.2$ cyc/oct.

The response $r[t]$ to this stimulus (middle panel)) exhibits both
linear and nonlinear aspects, as well as variability. According to the
linear model of Eq. (\ref{eq:re}), the response should be a pure $8$
Hz sinusoid, with amplitude and phase determined by $H[8,0.2]$.
Clearly, $r[t]$ (C: blue) is modulated at $8$ Hz, but it also contains
nonlinear components. The (Discrete) Fourier Transform $R[w]$
makes this explicit: In addition to a
prominent $8$ Hz component (in red), distortion products (in green)
with frequencies of $0$ Hz (the ``DC'' or average of $r[t]$ over $t$)
and $16$ Hz are plainly visible. Given the stimulus composition, these
distortion products betray the presence of
$2$\textsuperscript{nd}-order, and possibly
$0$\textsuperscript{th}-order (``spontaneous'' activity), nonlinearity
(both of which are even-order). With respect to the linear plus DC
description (red curve in inset panel), including the $16$ Hz distortion product
(black curve) better accounts for the sharpness and non-negative nature
of the response.

The remaining portion of the response looks like noise. It is the
manifestation of the period-to-period response variability. In the Fourier 
Transform of the response, it takes the form of a shallow
baseline of energy that extends over all frequencies. Note that the
square-root of the response variance (i.e., the standard error),
calculated via Eq. \ref{eq:rvar}, is similarly distributed over the
components of $R[w]$ (black curve).

\end{spacing}
\begin{spacing}{1}
\end{spacing}
\begin{spacing}{1.00}

The existence of response components due to nonlinearity and
variability does not necessarily imply that they interfere with the
STRF measurement. Since the stimulus consists of a single
spectrotemporal modulation frequency with a temporal component of $8$
Hz, only the $8$ Hz component of the response is correlated with the
stimulus, and $\tilde{\epsilon}$ in Eq. (\ref{eq:rc}) is zero. The
only source of error is the portion of the nonlinearity and
variability that happens to be manifest at $8$ Hz. The
reverse-correlation STRF (STRF$_{{DR}}$) can be assembled from results of 
consecutive presentations of dynamic ripples with different single spectrotemporal 
modulation frequencies. It is important to note that, due to time constraints, these point-by-point
measurements of $H$ were restricted to two
cross-sections, as indicated by the gray outlines in left panel (Bottom row). The 
full $H$ was then constructed from a normalized outer product
of these cross-sections \citep{dep1}.

\subsubsection{Temporally Orthogonal Ripple Combinations}
\label{sec:tc}

In contrast to the dynamic-ripple stimuli, the TORC stimuli
\citep{robust} can directly measure the entirety of the
$H$, because each stimulus is used to measure multiple
points at once. The stimuli are necessarily more complex, containing
six spectrotemporal modulation frequencies (Fourier components) each.
However, no two Fourier components in a given stimulus share the same
value of $|w|$ (they are temporally orthogonal; their temporal
correlation is zero); therefore, each spectrotemporal modulation
frequency in the stimulus will evoke a different temporal frequency in
the linear part of the response.


The dynamic spectrum of one TORC is shown in Figure \ref{fig:dr}B (top panel). It
is composed of six spectrotemporal modulation frequencies having the
same $\Omega$ of $0.2$ cyc/oct, but different $w$'s spanning the range
of $4$ to $24$ Hz. The associated response (inset in middle panel: blue) exhibits 
a complex modulation of the spike rate. The smoothed response, obtained
by discarding the frequencies above $24$ Hz, is
superimposed in red. A more accurate view of the linear part
of the response is also shown (in dashed black), which was obtained from the 
inverse-repeat procedure.
It is very similar to the response predicted by STRF$_{{DR}}$
(Figure \ref{fig:dr}A). The Fourier
Transform of the response confirms the strong presence of the $4$
to $24$ Hz components (in red) expected from the linear model.
However, with respect to the noise baseline, the response is weaker
than it was for the above dynamic-ripple stimulus.

In the reverse-correlation operation, the $4$ Hz response component is
orthogonal to all stimulus components besides the $4$ Hz component,
the $8$ Hz response component is correlated only with the $8$ Hz
stimulus component, and so on; $\tilde{\epsilon}$ is again zero. The STRF after
all stimuli are presented, are shown in bottom panels. It
bears a striking resemblance to STRF$_{{DR}}$,
despite the drop in both $SNR$ and $SNR_{cor}$. This indicates that
estimates of the neuron's STRF are robust to
changes in the spectrotemporal modulation frequency content of
stimuli.

\subsubsection{Spectrotemporally White Noise}
\label{sec:wn}

The spectrotemporally white
noise (STWN) is the most complex of the ripple-based stimuli; its $S$ contains all 
spectrotemporal
modulation frequencies (Fourier components) with equal amplitudes and
uniformly distributed phases. The typically poor efficacy of such
stimuli can be improved somewhat by limiting the $S$ to a
relevant range of spectral and temporal modulation frequencies
\citep{robust}. Figure \ref{fig:dr}C (top panel) shows the dynamic spectrum of one
such stimulus, which contained all $w$'s between $4$ and $24$ Hz and
all $\Omega$'s between $0$ and $1.4$ cyc/oct. Although the response
shown below (inset of middle panel) is quite a bit weaker than those observed in 
Figures
\ref{fig:dr}A and B, when smoothed (red) it is still
comparable to the linear predictions from both STRF$_{{DR}}$
(dashed black) and STRF$_{{TORC}}$ (dashed green); this is
despite the fact that the $4$ to $24$ Hz response
frequencies predicted by the linear model are barely distinct over the
noise baseline.


This reverse-correlation scenario differs from that of the other two
stimulus types. Each of the linear response frequencies is now the sum
effect of multiple Fourier components of the stimulus sharing the same
temporal modulation frequency. Every response frequency will in turn
be correlated with each of the stimulus components sharing the same
temporal modulation frequency. It is therefore not initially clear
which stimulus component caused what component of the response; all
points on the $H$ corresponding to a given $w$ cannot be
distinguished. This ambiguity manifests itself in the form of a large
$\tilde{\epsilon}$.

Because $\tilde{\epsilon}$ is dependent upon the (randomly assigned)
phases of the $S$, it has an incoherent structure that is
distributed over the entire measurement, and its strength can be
reduced by averaging the results from multiple stimuli with different
phases (or by using more finely spaced $w$'s, i.e., increasing the
base stimulus period $T$) \citep{robust}. This argument also applies
to the manifestations of variability and even-order nonlinearity
(some odd-order distortion products are however not dependent on the
phases of the stimulus frequencies \citep{victor}). The result
obtained after averaging the results from $30$ different stimuli is
shown in the bottom panels (approximately the same result would be obtained by
extending $T$ by a factor of $30$). Despite a further decrease in
$SNR$ and $SNR_{cor}$, its similarity to STRF$_{{DR}}$ (Figure
\ref{fig:dr}A) and STRF$_{{TORC}}$ (\ref{fig:dr}B) is impressive;
the STRF of the neuron has maintained its
form for more than an hour, over vastly different stimulus types.

\subsubsection{Application of the Singular-Value Decomposition}
\label{sec:svdill}

In this section, we demonstrate the use of the SVD for producing
approximations of the measurements of the STRF and $H$.
Such approximations represent an optimal trade-off between error
reduction and signal loss, provided the errors are evenly
distributed over the measurements \citep{stewart1, hansen}. The
proportion of signal lost is gauged by $\beta_{SVD}$ (see Methods).

\begin{figure}[!ht]
\begin{center}
\includegraphics{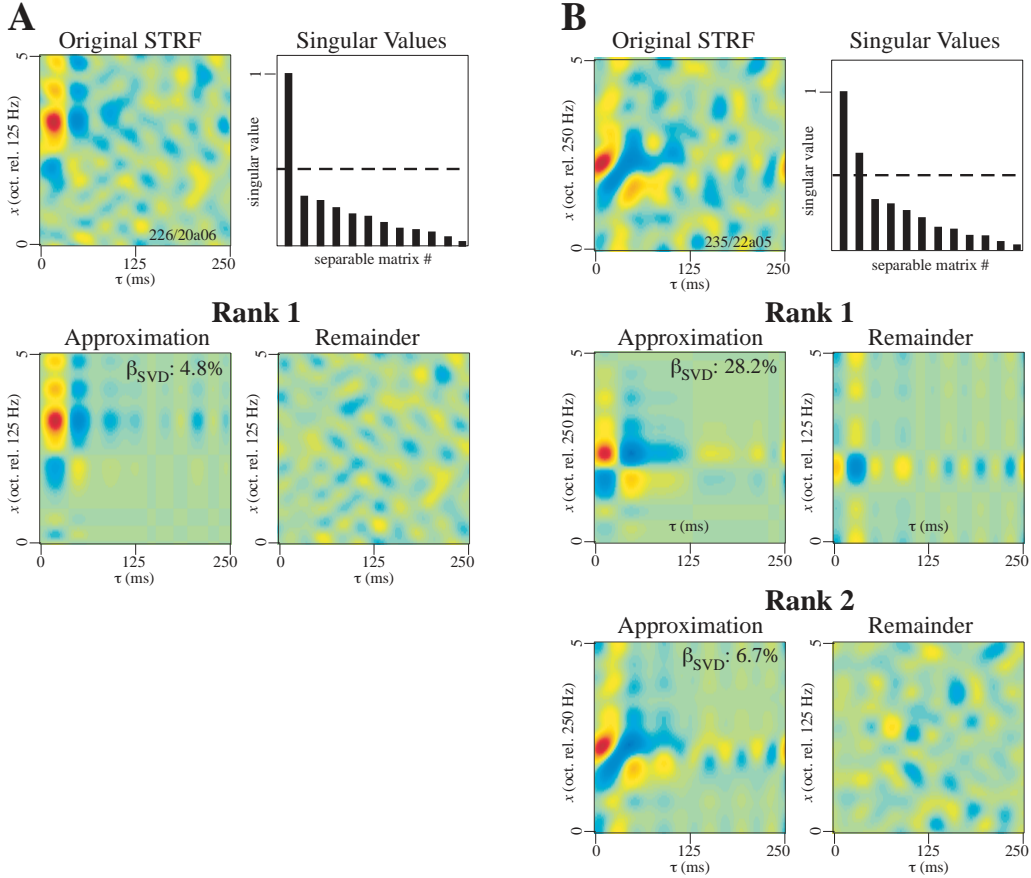}
\end{center}
\caption[SVD of the STRF]
{\footnotesize Approximating the STRF with the SVD. A: An STRF that looks
separable. (Top row) The original measurement, and the corresponding
singular values (bars) of the separable matrices of the SVD, and the
error-derived threshold (dashed line). (Bottom row) The {\em rank}-$1$
approximation and the remainder. B: An STRF that does not look separable.
(Top row) The original STRF and its corresponding singular values (bars)
and threshold (dashed line). (Middle row) The {\em rank}-$1$ approximation
and second-separable matrix (or remainder). (Bottom row) The {\em rank}-$2$
approximation, and the remainder.  A common color scale is shared by all
panels within A, and a different color scale is shared by all panels within
B.}

\label{fig:svd1}
\end{figure}

The SVD of the STRF$_{{TORC}}$ from Figure \ref{fig:dr}B is illustrated again 
in Figure
\ref{fig:svd1}A. The singular values of the first $12$
separable matrices from the SVD are shown (top row), along with the
error-derived threshold (see Methods) indicated by the dashed line.
The first singular value, corresponding to the separable rank 1 matrix (bottom 
row),
towers over the others, and alone exceeds the threshold. The STRF is
well described by this separable matrix, while the sum of the
remaining separable matrices, consists of unstructured
measurement errors. Indeed, $\beta_{SVD}=4.8\%$, indicating that more
than $95\%$ of the STRF power is captured by this {\em rank}-$1$
approximation. That is, in large part this STRF represents the product
of independent spectral and temporal integration.

In contrast, the SVD of a different neuron's STRF$_{{TORC}}$ is
shown in Figures \ref{fig:svd1}B. This STRF does not
look separable; for inputs at different tonotopic locations $x$, the
temporal integration by the neuron (in its network) is not related by
a simple scaling of the same function. In this case, the second
singular value (top row panel) also protrudes above the threshold, the {\em
rank}-$1$ approximation (middle panels) fails to describe the STRF's obliqueness,
and $\beta_{SVD}$ is high at $28.2\%$. After including the second
separable matrix (bottom row panels), the approximation is vastly
improved ($\beta_{SVD}=6.7\%$), and the remainder again chiefly
consists of unstructured errors.

\end{spacing}
\begin{spacing}{1}
\end{spacing}
\begin{spacing}{1.00}

\begin{figure}[!ht]
\begin{center}
\includegraphics{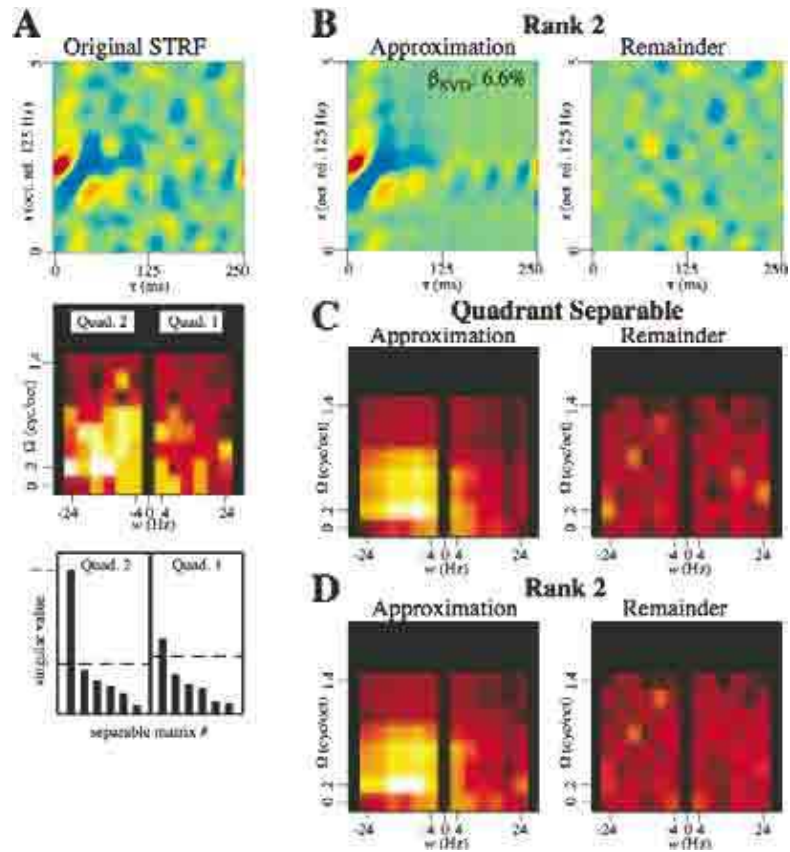}
\end{center}
\caption[SVD of the $H$] {\footnotesize Approximating the $H$ with the SVD.
A: (Top panel) The original STRF measurement and (Middle panel)the
corresponding $H$ magnitude (first two quadrants). (Bottom panel) The
singular values (bars) and thresholds (dashed lines) of the first two
quadrants of the $H$. B: The quadrant-separable approximation of the STRF
in A and the remainder. C:  The $H$ magnitude from the quadrant-separable
approximation, and the Fourier Transform of the remainder. D: The $H$
magnitude from the {\em rank}-$2$ approximation (from Figure
\ref{fig:svd1}), and the Fourier Transform of the remainder (from
\ref{fig:svd1}) A common color scale is shared by B and the top panel of A
. A different color scale is shared by C, D, and the middle panel of A.
}
\label{fig:svd2}
\end{figure}

The SVD can alternatively be applied to the $H$. While
the SVD of the {\em full} $H$ yields an approximation
identical to that of the STRF, applying the SVD separately to each of
the {\em quadrants} of the $H$ will generally produce a
different approximation. This procedure is of interest chiefly because
previous studies (using dynamic-ripple stimulation) have suggested
that the $H$'s of AI neurons are well described as being
{\em quadrant-separable} \citep{kowalski2, dep1}, implying that the
SVD of each quadrant of the $H$ should yield at most one
separable matrix of significance. Therefore, if the STRF is not
separable, it could be advantageous (in terms of error reduction) to
approximate the STRF in this manner. This principle is examined in
Figure \ref{fig:svd2}, using the non-separable STRF from Figure
\ref{fig:svd1}B. The SVD of each of the upper two quadrants of the
$H$ shown in \ref{fig:svd2}A (middle panel) yields the two sets of
singular values (bottom panel). In each quadrant, only the first singular value
is pronounced and exceeds the threshold. This indication that the
quadrants are indeed separable is supported upon comparison of the
original STRF (top panel) with the quadrant-separable approximation (for
which $\beta_{SVD}=6.6\%$) and the remainder, shown in B. Intriguingly, the result 
is markedly similar to the {\em
rank}-$2$ approximation of the STRF from Figure \ref{fig:svd1}B. By
implication, the $H$ from the {\em rank}-$2$
approximation (shown in D) is very similar that from the
quadrant-separable approximation (in C). The Fourier Transforms of the
corresponding remainders are also very similar.

\end{spacing}
\begin{spacing}{1}
\end{spacing}
\begin{spacing}{1.00}

In summary, we have demonstrated the use of the SVD for producing
relatively error-free approximations of the STRF or $H$
measurements. Later, in Section \ref{sec:svdres}, we will examine how
well these three types of approximations --- the {\em rank}-$1$, {\em
rank}-$2$, and quadrant-separable approximations --- apply to the
whole of the neuronal population, as a function of the error level and
the type of stimulation.

\end{spacing}
\begin{spacing}{1}
\subsection{Direct Comparisons of STRFs Measured with Different
Stimulus Types}
\label{sec:ccs}
\end{spacing}
\begin{spacing}{1.00}

In $45$ out of $308$ neurons whose STRFs we measured, we obtained
multiple STRF measurements using two or all three stimulus types. The
resemblance between the first $125$ ms of each pair of measurements
was quantified by the correlation coefficient (see Methods), which was
computed under four conditions: for the raw (pre-SVD) measurements,
and for the quadrant-separable, {\em rank}-$2$, and {\em rank}-$1$
approximations of the measurements.

\begin{figure}[!hp]
\begin{center}
\includegraphics[width=.94\linewidth]{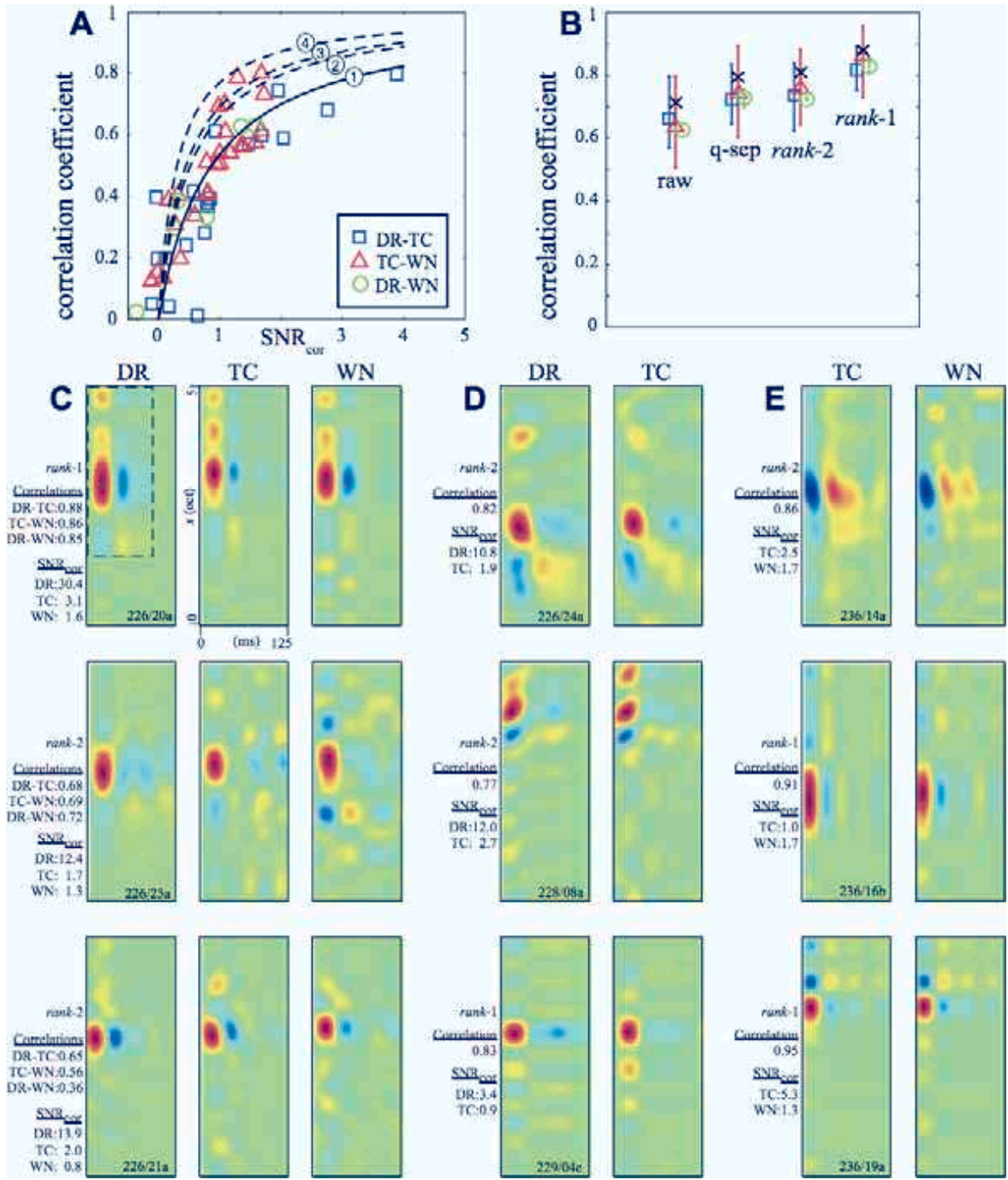}
\end{center}
\caption[Direct STRF Comparisons]
{\footnotesize Similarity between STRFs measured with different
stimulus types: Dynamic ripple (DR), TORC (TC) and STWN (WN). DR-TC,
e.g., denotes comparisons between DR and TORC STRFs.
Correlation coefficients were computed between the original (raw)
measurements, and between the quadrant-separable (q-sep), {\em
rank}-$2$ and {\em rank}-$1$ approximations of each measurement. A:
Correlation coefficients plotted versus minimum $SNR_{cor}$ of
the two original measurements. Squares, triangles, and circles
correspond to the raw comparisons; different symbols correspond to the
different pairs of stimulus types compared (see legend). Curve 1
(solid black) is the relationship expected from two identical STRFs
with independent errors. Curves 2, 3, and 4 (dashed curves) are fits
to the correlation coefficients obtained from the quadrant-separable,
{\em rank}-$2$, and {\em rank}-$1$ approximations of each measurement,
respectively (see text). B: The complete range (vertical lines) and
the average (squares, triangles, and circles) of the correlation
coefficients are shown for all comparisons where the minimum
$SNR_{cor}$ was above $1$. Also shown (black x's) are the average
correlation coefficients, for all pairs of stimulus comparisons
combined, obtained when the comparison is further limited to the
half-sized rectangular region of the STRF containing the most power
(see, e.g., dashed box on the top left STRF of Column C). Columns
C, D, E: In each row, the STRF of the same neuron measured with
different stimulus types are shown side by side. Shown are either the
{\em rank}-$1$ or {\em rank}-$2$ approximations of the STRFs,
depending on what was optimal for the measurement with the highest
$SNR_{cor}$. To the left are correlation coefficients obtained
from each pair of comparisons and the $SNR_{cor}$ of the original
measurements.}
\label{fig:ccs}
\end{figure}

The correlation coefficients from the raw comparisons are plotted in
Figure \ref{fig:ccs}A versus the limiting (minimum) $SNR_{cor}$ of the
two measurements. The squares, triangles, and circles correspond to
the three possible pairs of stimulus types compared. The trends
followed by all stimulus comparisons are similar. When $SNR_{cor}$ is
above $1$, the correlation coefficients are high and are weakly
affected by $SNR_{cor}$. The correlation coefficients are only small
when $SNR_{cor}$ is small; as $SNR_{cor}$ descends to $0$, so do the
correlation coefficients. This mirrors the relationship expected from
two identical STRFs that are corrupted by independent errors, as
indicated by the solid-black Curve 1. In other words, it the
relationship produced when the STRFs of
the system, summarized by the (error-free) STRF, is impervious to
changes in stimulus type, but the STRF measurement is error-prone.

Since the SVD approximations act to reduce errors, they should result
in higher correlation coefficients, provided the STRF measurements
have similar signal components. These properties are evident in the
three dashed curves in \ref{fig:ccs}A, which summarize the correlation
coefficients obtained from the quadrant-separable (Curve 2), {\em
rank}-$2$ (Curve 3), and {\em rank}-$1$ (Curve 4) approximations of
each pair of measurements (the data points are not shown, for
clarity). The curves fit the combined data from all three types of
stimulus comparisons. The fits were produced by modeling the error
reduction as a multiplicative gain $g$ in $SNR_{cor}$ (see Methods).
The values of $g$ used for Curves $2$--$4$ are $1.7$, $1.9$, and
$2.9$, respectively; these values minimized the number of data points
deviating more than $0.1$ units away from the curves (providing the
most visually pleasing fits).

\end{spacing}
\begin{spacing}{0.85}
\end{spacing}
\begin{spacing}{1.00}

For all data points exceeding the critical $SNR_{cor}=1$ level, Figure
\ref{fig:ccs}B shows the complete range and the average of the
correlation coefficients. Again, similar results are obtained no
matter which two stimulus types are compared. For the raw
measurements, correlation coefficients fall between $0.5$ and $0.8$,
with an average of $0.64$. The average rises to $0.73$ and $0.75$ for
the quadrant-separable and {\em rank}-$2$ approximations,
respectively. For the {\em rank}-$1$ approximations, the correlation
is $0.85$ on average, is as high as $0.97$, and does not fall below
$0.74$. The average correlations are still higher ($0.71$, $0.78$,
$0.80$, and $0.88$, respectively) when the comparisons are further
restricted to the half-sized rectangular region containing the most
power (e.g., the dashed box in the top row of \ref{fig:ccs}C), as
indicated by the x's. Least affected are the {\em rank}-$1$
comparisons, suggesting that they are already relatively error free.
Note that these values far surpass those typically produced by
comparing the STRFs of different neurons; for example, if the {\em
rank}-$1$ approximation of a neuron's STRF$_{{TORC}}$ was
compared to the {\em rank}-$1$ approximation of the subsequent
neuron's STRF$_{{STWN}}$, the average correlation was $0.03$.

Some visual comparisons of STRF measurements are available in Columns
C through E of Figure \ref{fig:ccs}. For each comparison, either the
{\em rank}-$1$ or {\em rank}-$2$ approximations are shown, depending
on what was optimal for the STRF with the highest $SNR_{cor}$. In C
are results from three neurons that were tested with all three
stimulus types. A typical {\em rank}-$1$ result is shown in the top
row. The STRFs match in many details, including the suppressive areas
and the multiple excitatory areas. In the middle row is a {\em
rank}-$2$ example with somewhat lower-than-average correlation
coefficients. While some features match well across stimuli, there is
an increase in background fluctuations between STRF$_{{DR}}$
and STRF$_{{STWN}}$ that limits the comparisons. The {\em
rank}-$1$ approximations may have been more appropriate here (and
these yielded correlation coefficients over $0.8$). In the bottom row
is an unusual {\em rank}-$2$ example, where the $STRF$ peak shifts to
a higher frequency, thus diminishing the correlation coefficients.
However, $SNR_{cor}$ of the STRF$_{{STWN}}$ was only $0.8$,
so it is difficult to make definite claims about its structure.
Results from additional neurons that were tested with two of the three
stimulus types are provided in D and E. Overall, a wide variety of
STRFs shapes, including unusual ``offset'' types (E, top row), are
well preserved across stimulus type. To be sure, there is much less
variation in STRF shape across stimulus type than there is across
neurons.

In summary, both visual and quantitative comparisons reveal a close
resemblance between the STRFs measured with different stimulus types.
The resemblance predictably increases as the limiting $SNR_{cor}$ of
the measurements increases; similarly, using the SVD to reduce the
error level only serves to increase their resemblance. The highest
correlation coefficients result from the {\em rank}-$1$
approximations, indicating that they are the most error-tolerant.
Similar results are obtained no matter which of the three possible
pairs of stimulus types are compared. By the same token, a wide
variety of STRFs are observed across neurons. 

Together, these
observations indicate that linear spectrotemporal processing is a
robust property of AI that takes diverse forms in individual neurons.

\subsection{The Sources and Stimulus Dependence of Measurement Error}
\label{sec:err}

In Section \ref{sec:ccs}, it was shown that the signal component of
the STRF measurement, seen through the corrective lens of the SVD, is not
crucially dependent on the stimulus type. Instead, the ability of the
SVD to separate this signal from the measurement errors is crucially
dependent on $SNR_{cor}$, which may depend on the stimulus type. In this
section, we examine the sources contributing to $SNR_{cor}$ and their
stimulus dependence.

\subsubsection{Systematic Error}
\label{sec:syserr}

\begin{figure}[!ht]
\begin{center}
\includegraphics{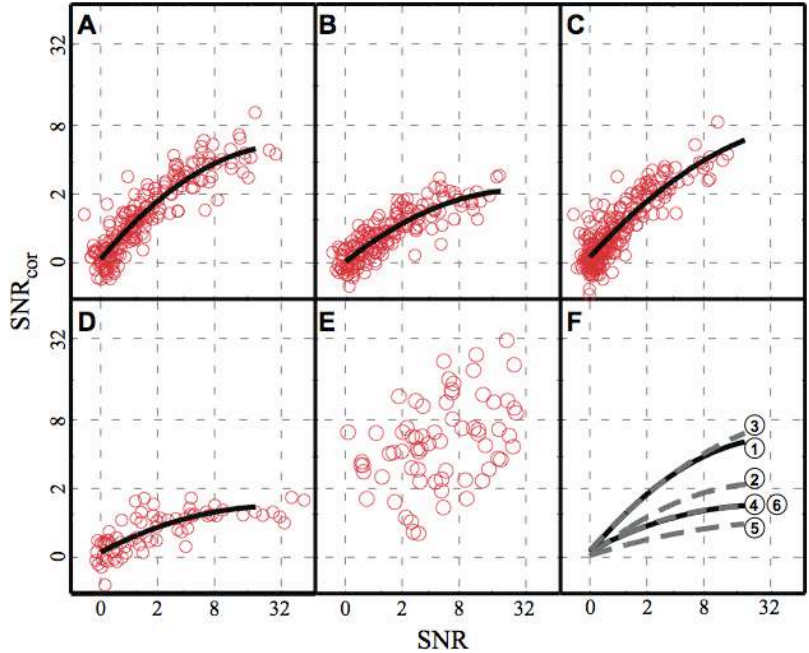}
\end{center}
\caption[$SNR$ versus $SNR_{cor}$]
{\footnotesize The relationship between $SNR$ and $SNR_{cor}$ across
all measurements for each stimulus type, and second-degree polynomial
fits (black curves) when appropriate. The level of saturation of these
curves indicates the relative levels of systematic errors in the
measurements. A: TORC. B: TORC without inverse-repeat, thus retaining
systematic errors due to even-order nonlinearities. C: TORC control,
discarding half of the stimulus presentations, D: STWN. E: Dynamic
ripple. F: Comparison of polynomial fits: Curves 1--4 are from Figures
A--D. Curve 5: STWN, discarding half of the stimuli, thus increasing
systematic errors induced by the stimulus ($\tilde{\epsilon}$). Curve
6: STWN control, discarding half of the stimulus presentations.}
\label{fig:snreff}
\end{figure}

The capacity of systematic errors to limit the quality of the
measurements is evident in the relationship between $SNR$ and
$SNR_{cor}$. This relationship, observed over all measurements for
each stimulus type, is plotted in Figure \ref{fig:snreff} (with
second-degree polynomial fits where appropriate). For both the TORC
(A; F, Curve 1) and the STWN (D; F, Curve 4) measurements, $SNR_{cor}$
shows a clear saturating characteristic as $SNR$ increases. Recall
that $SNR_{cor}$ incorporates both the non-systematic and systematic
errors, while the SNR incorporates only the non-systematic errors.
Therefore, as the measurements become more reliable (SNR increases),
the saturation of $SNR_{cor}$ evinces the systematic error that
dominates when the non-systematic errors are sufficiently small. The
relative significance of the systematic error component is revealed in
the level to which $SNR_{cor}$ is limited in the high SNR
measurements.

\end{spacing}
\begin{spacing}{1}
\end{spacing}
\begin{spacing}{1.00}

Recall that for the TORC measurements (A; F, Curve 1), the
inverse-repeat method was employed in order to remove systematic
errors due to even-order nonlinearities. Therefore, the saturation of
$SNR_{cor}$ in the TORC measurements should be worsened if the
inverse-repeat method is not used. Indeed, bypassing the
inverse-repeat method did further limit $SNR_{cor}$ (B; F, Curve 2),
by a factor of about $2.5$. Note that this is not simply a side effect
of $SNR$ reductions caused by discarding half of the data, for it is
not observed if half of the stimulus presentations are discarded but
inverse-repeat is still employed (C; F, Curve 3).

In the STWN measurements (D; F, Curve 4), the systematic errors are
much more severe than in the TORC measurements; the limiting value of
$SNR_{cor}$ is at least $4$ times lower, and so $SNR_{cor}$ is much
less likely to exceed usable values. $SNR_{cor}$ is also less variable
across the high-SNR measurements; when the measurements are reliable,
which is fairly often, $SNR_{cor}$ reliably reaches its limited
potential. This potential is further cut in half by discarding half of
the stimuli (F, Curve 5), but not by discarding half of the
presentations of each stimulus (Curve 6). In sum, these observations
suggest that the errors are dominated by the nonideality of the STWN
stimuli (i.e., $\tilde{\epsilon}$), to which all neurons were exposed.
Our simulations also supported this view. Therefore, at least $4$
times as many STWN stimuli would have to be used in order to raise the
$SNR_{cor}$ potential to the level of the TORC method.

Finally, note that the relationship between $SNR$ and $SNR_{cor}$ is
less clearly defined in the dynamic-ripple measurements (E) (although
both $SNR_{cor}$ and $SNR$ often surpass the values achieved by the
other two stimulus types). In our experience, this is largely because
the errors are not uniformly distributed over the dynamic-ripple STRFs
\citep{dep1}, due to the outer-product operation in the construction of
the $H$. As a result, $SNR_{cor}$ is a less reliable
gauge of the overall error level in the dynamic-ripple measurements.

\subsubsection{Non-Systematic Error}
\label{sec:varerr}

In Section \ref{sec:syserr}, it was shown how the potential accuracy
of the STRF measurements is limited by the level of systematic error,
which depended on the stimulation method. However, if a method is to
achieve a given level of accuracy within its potential, it is evident
in Figure \ref{fig:snreff} that the SNR (which reflects the level of
non-systematic error) must be at least minimally adequate. In this
section, we explore how the SNR is determined from the interplay
between the stimulus, the STRF, and the neuronal response.

To set the stage, recall from Eq. (\ref{eq:tfest}) that a single
stimulus-response pair results in the measurement of a set of one or
more points on $H[w,\Omega]$, which is given by the
spectrotemporal modulation frequencies content of the stimulus. By Eq.
(\ref{eq:tfvar}), the variance of each point $(w,\Omega)$ is a fixed
proportion, namely $1/a^2$, of the variance of the response's Fourier
Transform at the corresponding (temporal) frequency $w$ ($a^2$ is the
power of each of the spectrotemporal modulation frequencies in the
stimulus). Now, consider the whole of the $H$
measurement, built stimulus-by-stimulus. To simplify matters, we will first 
consider the
situation in which every point of the measurement has resulted from a
{\em single} stimulus-response pair --- that is, prior to the TORC
inverse-repeat procedure, the STWN phase-averaging procedure, or the
dynamic-ripple outer-product operation. In that case, to find the
variance of any point on the $H$, one needs only to find
the variance of the appropriate response at the appropriate frequency,
and weight it by $1/a^2$. Consequently, the average variance of the
entire $H$ (and STRF) measurement,
$\left<\sigma^2\right>$, is simply $1/a^2$ times the average variance
at all of the relevant frequencies of all responses. The SNR is then
the ratio of $P$ (the STRF signal power) to this number.

\begin{figure}[!ht]
\begin{center}
\includegraphics{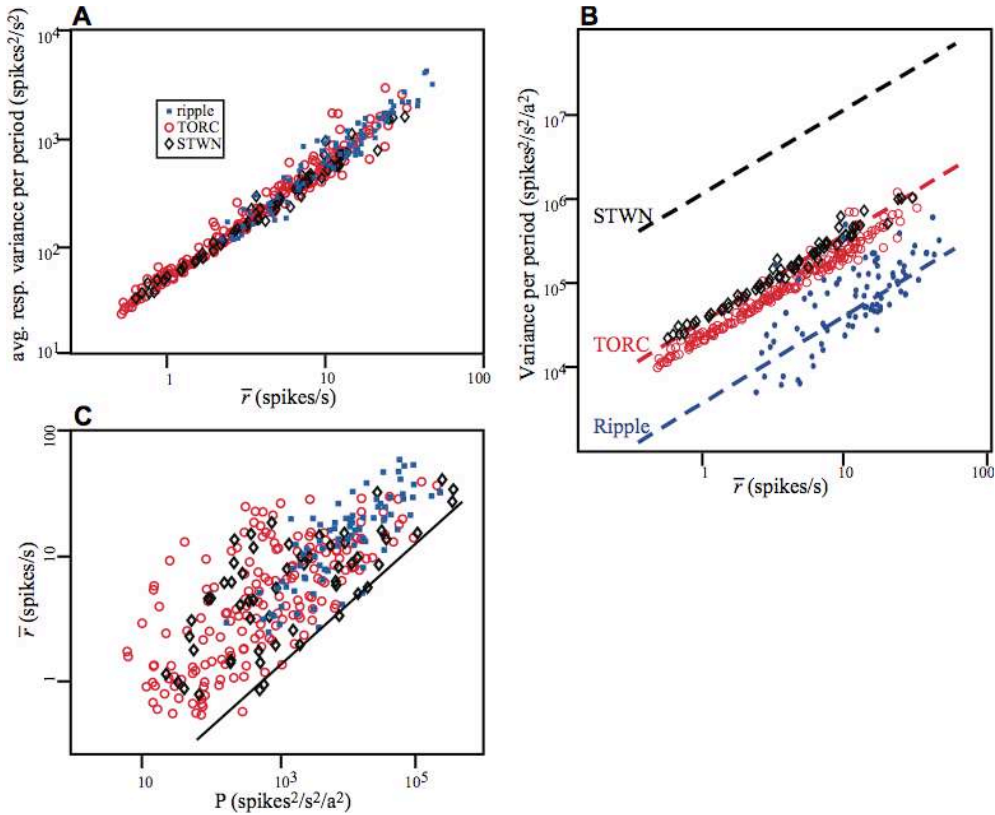}
\caption[$SNR$]
{\footnotesize The sources and stimulus dependence of $SNR$, for the
dynamic ripple stimuli (blue dots), TORCs (red circles), and STWN
(black diamonds). A: The linear relationship between the average spike
rate $\bar{r}$ and the average variance of the response's Fourier
Transform. $\bar{r}$ is averaged over all responses. The variance is
averaged over all responses, but only those temporal frequencies of
each response relevant to the $H$ measurement (where the
corresponding $S$ magnitude was nonzero, e.g., $w=4,8,É,24$ Hz). The variance is
scaled by $n$ (the number of stimulus periods) to correct for
differences $n$ across the measurements, and thus represents the
variance of a single response period. B: (dashed lines) The expected
relationships between $\bar{r}$ and $\left<\sigma^2\right>$ (scaled by
$n$) in the case where each point of the $H$'s is
obtained from a single stimulus-response pair. The actual
relationships observed (plotted points) differ from the dashed lines
by an amount predicted by the number of stimulus-response pairs whose
results are averaged to obtain the final $H$ (see text).
C: The lower bound of $\bar{r}$ is proportional to the square-root of
the STRF signal power $P$ (the diagonal line's slope is $1/2$). The
square-root law is expected from a linear-plus-rectification response
model, but the scatter in $\bar{r}$ suggests additional sources of
variability.}
\label{fig:snr}
\end{center}
\end{figure}

What determines the variance of a response's Fourier Transform? Two
observations lead to a simple answer. First, as middle panels in Figures
\ref{fig:dr}A, \ref{fig:dr}B, and \ref{fig:dr}C typify, the variance of
$R[w]$ is nearly frequency-invariant (deviations from this could reflect
refractoriness, burstiness, or oscillations in the response \citep{bair}).
Therefore, the average variance over the relevant frequencies is closely
related to the average variance over {\em all} frequencies. Now, the
average variance over all frequencies equals the average variance over all
times \citep{papoulis, oppenheim}, which ties in the second observation:
The variance of $r[t]$ is proportional to $r[t]/n$ (where $n$ is the number
of stimulus periods). This originates from a linear relationship between
the sample mean and the sample variance of the binned spike train responses
($y[t]$), which is a widely reported observation \citep{shadlen}.
Consequently, the average variance over time is proportional to the average
spike rate over time, $\bar{r}$.  So finally, all else being equal across
stimuli, $\bar{r}$ (over all responses) can be treated as the lone variable
determining the average variance of the responses over the relevant
frequencies. The relationship observed across all STRF measurements is
shown in Figure \ref{fig:snr}A, where the variance has been transformed
into the variance of a single response period by multiplying by $n$ (thus
correcting for differences in $n$ across measurements). The trend across
all neurons is indeed linear (on this log-log plot, the slopes of the
linear fits to the data were very close to $1$), and is only weakly
influenced by stimulus type.

\end{spacing}
\begin{spacing}{1}
\end{spacing}
\begin{spacing}{1.00}

In contrast, the choice of stimulus type effects order-of-magnitude
differences in $a^2$ (due to differences in the number of
spectrotemporal modulation frequencies per stimulus; recall Figure
\ref{fig:as}). This in turn strongly effects the STRF variance
$\left<\sigma^2\right>$ for a given average spike rate $\bar{r}$.
Given the relationship observed between $\bar{r}$ and average response
variance in \ref{fig:snr}A, the predicted relationship between
$\bar{r}$ and $\left<\sigma^2\right>$ (again scaled by $n$) for each
of stimulus type is indicated by the dashed lines in B. Note, however,
that for a given neuron, the actual effect of stimulus type on
$\left<\sigma^2\right>$ depends on how $\bar{r}$ is also affected.
Curiously, we have seen little evidence for a significant effect of
stimulus type on $\bar{r}$. From one type to the next, up to
factor-of-two increases or reductions in $\bar{r}$ were typical, but
this variation is not systematic and is small compared that of $a^2$.

The actual relationship between the average spike rate and the STRF
variance observed across all STRF measurements is indicated by the
data points plotted in B. The discrepancies between these trends and
the dashed lines, where they exist, are easily explained by the fact
that every point of the actual $H$ measurements is {\em
not} the result of {\em just one} stimulus-response pair, as we have
so far assumed. For the STWN stimuli, $H[w,\Omega]$ was the
average result from $30$ stimulus-response pairs; therefore, its
actual variance $\left<\sigma^2\right>$ (black diamonds) was lower
than the black (upper-most) dashed line by a factor of $30$. This
largely compensated for the difference in $a^2$ between the STWN and
TORC stimuli. Similarly, the inverse-repeat method effectively
averages the results from two sets of stimuli, and so the
$\left<\sigma^2\right>$ of the final TORC result (red circles), was
cut in half with respect to the red (middle) dashed line. Finally, we
observed that the $\left<\sigma^2\right>$ of the final dynamic-ripple
$H$ (blue dots), each point of which results from the
normalized product of two individual measurements, was typically
similar to that of the measured cross-sections alone. Therefore, its
relation to $\bar{r}$ was similar to the black (lower-most) dashed
line, albeit with quite a bit of scatter. Overall, these properties
conspired to produce $SNR$'s that were, on average, a factor of $5$
lower in the TORC measurements than in the dynamic-ripple
measurements, and an additional factor of $2$ lower in the STWN
measurements.

For each stimulus type, the average spike rate $\bar{r}$ observed
across neurons ranged over roughly two orders of magnitude. Figure
\ref{fig:snr}C shows that the value of $\bar{r}$ is partially
predicted by the STRF power $P$, in that $\bar{r}$, and more strictly
its lower bound, tends to grow by the {\em square-root} of $P$ (the
black line on this log-log plot has a slope of $1/2$). A square-root
relationship is expected from the linear response model followed by
rectification: Generally speaking, STRFs (and $H$'s)
with higher magnitudes result in spike rates with proportionally
stronger modulations, which, since the spike rate must be positive,
result in proportionally higher $\bar{r}$'s; meanwhile, $P$ grows as
the {\em square} of the STRF magnitudes. Since $\bar{r}$ translates
{\em linearly} into variance, this implies that STRFs with higher
average power $P$, although associated with higher absolute levels of
variability, have the potential to achieve higher SNRs; and this
potential is realized in those neurons with the lowest $\bar{r}$
allowed for a given $P$. Note that the data from all stimulus types
overlap, reinforcing the idea that $\bar{r}$ is not significantly
affected by stimulus type.

In summary, the ingredients of $SNR$ are of two largely independent
varieties: properties of the stimulus and properties of the auditory
system. The key stimulus properties boil down to the power in each
spectrotemporal modulation frequency $a^2$, to which the SNR is
inversely proportional, and the number of stimulus-response pairs used
to measure each point of the $H$ (including $n$, the
number of periods of each stimulus), to which $SNR$ is proportional.
The system properties reduce to the STRF power $P$ and the average
spike rate $\bar{r}$, to which the SNR is proportional and inversely
proportional, respectively. Furthermore, $\bar{r}$ can be seen as the
sum of two positive-valued components. One is proportional to the
square-root of $P$, as predicted by a linear-plus-rectification
response model. The other not obviously related to the STRF, and
represents an additional source of variability that varies in strength
from neuron to neuron. The net result is that an increase in $P$
serves to increase the SNR, while, for a given $P$, an increase
$\bar{r}$ counteracts this effect.

\end{spacing}
\begin{spacing}{1}
\subsection{Sufficiency and Error Dependence of the SVD-Based
Approximations}
\label{sec:svdres}
\end{spacing}
\begin{spacing}{1.00}

In Section \ref{sec:ccs}, the SVD approximations of STRFs measured
with different stimulus types were found to be highly similar when
$SNR_{cor}$ (which reflects the level of measurement error) was
adequate in both measurements. The stimulus dependence of $SNR_{cor}$
was then analyzed in detail in Section \ref{sec:err}. In this section,
we further examine how the SVD approximations are influenced by
$SNR_{cor}$. Primarily, we are concerned with the extent to which
measurement errors may prevent the SVD from resolving features of the
``true'' (error-free) STRF.

\begin{figure}[!ht]
\begin{center}
\begin{minipage}[b]{0.57\linewidth}
\raisebox{-1.0cm}{\includegraphics[width=\linewidth]{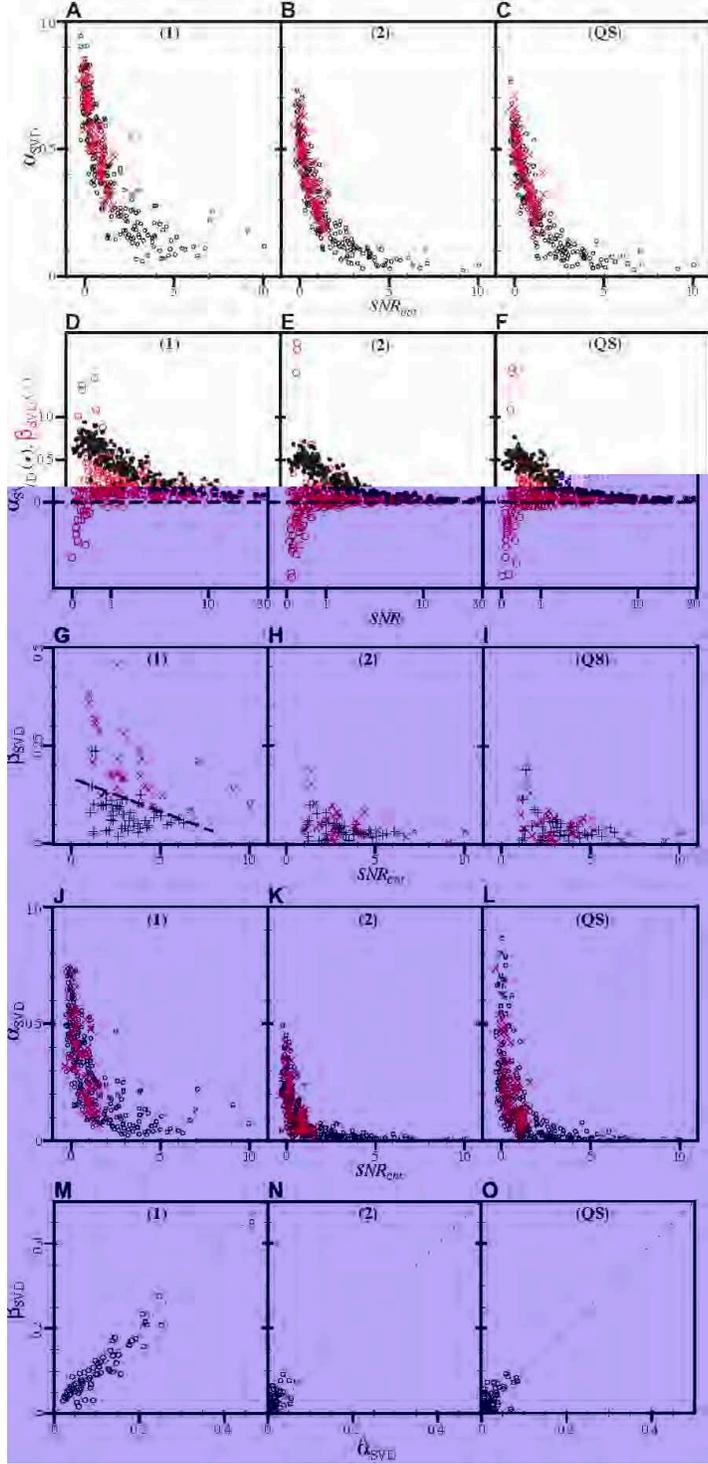}}
\end{minipage}\hfill
\begin{minipage}[b]{0.42\linewidth}
\caption{\footnotesize Sufficiency of the SVD approximations as a
function of the error level. A--C: $\alpha_{SVD}$, the proportion of
the STRF measurement's power lost from the SVD approximations, for all
TORC (black o's) and STWN (red x's) measurements. D--F: $\beta_{SVD}$
(red o's) and $\alpha_{SVD}$ (back dots) versus $SNR$ for all TORC
measurements. $\beta_{SVD}$ estimates the proportion of the {\em
systematic part} of the STRF measurement relegated to the SVD
remainder, and therefore better reflects the true STRF's structure. At
very low SNRs, $\beta_{SVD}$ is unstable (some points lay beyond the
axis limits). G--I: $\beta_{SVD}$ versus $SNR_{cor}$ for all TORC
measurements with $SNR$ above $1.5$. Black +'s and red x's denote
those measurements optimally approximated by {\em rank}-$1$
(separable) and {\em rank}-$2$ (non-separable) matrices, respectively.
With $\beta_{SVD}^{(1)}$ (G) typically as high as $25\%$, many STRFs
are not well described by the {\em rank}-$1$ approximations. In
contrast, $\beta_{SVD}^{(2)}$ (H) and $\beta_{SVD}^{(QS)}$ (I) are
typically well below $10\%$, indicating that all STRFs are well
described by {\em both} the {\em rank}-$2$ and quadrant-separable
approximations. The unusually high $\beta_{SVD}$'s at the lowest
$SNR_{cor}$'s indicates that the SVD is unable to resolve the
structure of some non-separable STRFs with high error levels. J--L:
$\hat{\alpha}_{SVD}$, computed as $\alpha_{SVD}$ but from the
quadrant-separable (J, K) and the {\em rank}-$2$ (L) approximations of
the TORC (black o's) and STWN (red x's) measurements. M--O: As
expected, the $\hat{\alpha}_{SVD}$'s are well matched to the
corresponding to $\beta_{SVD}$'s in those TORC measurements with
$SNR_{cor}$ above $2$.}
\label{fig:svd3}
\end{minipage}
\end{center}
\end{figure}

For this purpose, it would be useful to know the proportion of the
true STRF's power lost from an SVD approximation of the measurement.
Unfortunately, in the presence of measurement error, this quantity is
not precisely knowable. One way to estimate it is to compute the
proportion of the STRF {\em measurement's} power lost from an SVD
approximation, which we call $\alpha_{SVD}$ \citep{dep1}. In total, we
will consider $\alpha_{SVD}^{(1)}$, $\alpha_{SVD}^{(2)}$, and
$\alpha_{SVD}^{(QS)}$, which speak to the sufficiency of the {\em
rank}-$1$, {\em rank}-$2$, and quadrant-separable approximations,
respectively. One obvious disadvantage of $\alpha_{SVD}$ is that it is
inflated in the presence of measurement errors (which comprise much of
the measurement's lost power). This is evident in Figures
\ref{fig:svd3}A through C, where $\alpha_{SVD}^{(1)}$ (A),
$\alpha_{SVD}^{(2)}$ (B), and $\alpha_{SVD}^{(QS)}$ (C) are plotted
versus $SNR_{cor}$ for all TORC and STWN STRFs (recall that
$SNR_{cor}$ is unreliable for the dynamic-ripple STRFs). The influence
of $SNR_{cor}$ on $\alpha_{SVD}$ clearly persists up to high
$SNR_{cor}$'s.

\end{spacing}
\begin{spacing}{1}
\end{spacing}
\begin{spacing}{1.00}

We reduced the dependence of $\alpha_{SVD}$ on the error level by
removing the effect of the non-systematic errors (see Methods). The
improved measure, $\beta_{SVD}$ is a more accurate gauge of the
proportion of lost STRF power, especially when the systematic errors
are small (e.g., in the TORC measurements). In theory, $\beta_{SVD}$
should be more tolerant than $\alpha_{SVD}$ to changes in SNR, and
$\alpha_{SVD}$ should converge down to $\beta_{SVD}$ with increasing
$SNR$. These properties are verified in Figures \ref{fig:svd3}D
through F, where $\beta_{SVD}$ (red circles) and $\alpha_{SVD}$ (back
dots) are plotted versus $SNR$ for the TORC measurements (the only
caveat is that at very low SNRs, $\beta_{SVD}$ becomes unstable). It
is concluded (with additional support from our simulations) that at
moderate to high $SNR$s, the effect of non-systematic error is
accurately removed in the computation of $\beta_{SVD}$. Therefore,
$\beta_{SVD}$ estimates the proportion of the {\em systematic part} of
the STRF measurement relegated to the SVD remainder, and better
reflects the true STRF's structure. To be conservative, we will
consider $\beta_{SVD}$ only in those measurements with $SNR$'s over
$1.5$.

The relationship between $\beta_{SVD}$ and $SNR_{cor}$ for the $82$
TORC measurements meeting this criterion is plotted in Figures
\ref{fig:svd3}G through I. The blue +'s and red x's denote the $50$
and $31$ measurements optimally approximated by {\em rank}-$1$ and
{\em rank}-$2$ matrices, respectively (the lone {\em rank}-$3$
approximation is not shown). At moderate to high $SNR_{cor}$'s (e.g.,
above $2$), the $\beta_{SVD}$ distributions are only weakly dependent
on $SNR_{cor}$. In other words, the SVD approximations are only weakly
affected by measurement errors, and therefore $\beta_{SVD}$ should
more accurately reflect the structure of the true STRF. Therefore, the
typical range of $\beta_{SVD} ^{(1)}$ (\ref{fig:svd3}G), roughly from
$3\%$ to $25\%$, indicates many STRFs are poorly described by {\em
rank}-$1$ approximations. It is reassuring that the lower and upper
portions of this range are dominated by the measurements optimally
approximated by {\em rank}-$1$ and {\em rank}-$2$ matrices,
respectively. However, the boundary between the two populations
progressively shifts from about $5\%$ at the highest $SNR_{cor}$ to
nearly $15\%$ at the lowest $SNR_{cor}$. This reflects the fact that
the optimal trade-off between error reduction and signal loss afforded
by the SVD approximations gets worse as $SNR_{cor}$ decreases; at
higher error levels, the true STRF must be further from being {\em
rank}-$1$ before the second separable matrix of the SVD becomes
dominantly signal and is included in the approximation.

Over this same range of suitably high $SNR_{cor}$'s,
$\beta_{SVD}^{(2)}$ (H) and $\beta_{SVD}^{(QS)}$ (I) are universally
bound below $10\%$, with averages of $3.1\%$ and $3.6\%$,
respectively. That is, the true STRFs are almost completely contained
within {\em both} the {\em rank}-$2$ and quadrant-separable
approximations of TORC measurements with suitably low error levels.
Indeed, as was illustrated in Section \ref{sec:svdill}, the two
approximations were usually very similar.

When $SNR_{cor}$ is low, a handful of measurements have conspicuously
high values of $\beta_{SVD}^{(1)}$ (G), $\beta_{SVD}^{(2)}$ (H), or
$\beta_{SVD}^{(QS)}$ (I). There are three plausible reasons for this:
(1) The systematic errors in these measurements are unusually large
(thus inflating $\beta_{SVD}$); (2) The true STRFs are actually poorly
described by these SVD approximations, and coincidentally the
measurements have a high error level; (3) Because of the high error
level, the SVD of these STRFs shapes is being disrupted, and more STRF
power is being lost than otherwise would be. We favor the last reason,
since (despite the error level) most of these STRFs appear to have
non-separable shapes. Such STRFs are are also found at higher
$SNR_{cor}$'s, but these high values of $\beta_{SVD}^{(2)}$ and
$\beta_{SVD}^{(QS)}$ are not found at higher $SNR_{cor}$'s.

Although they are needed to fully describe many STRFs, the trade-off
to using the {\em rank}-$2$ or quadrant-separable approximations
instead of the {\em rank}-$1$ approximations is that they retain a
higher proportion of the measurement error. This was earlier indicated
in Figures \ref{fig:ccs}A and B. Similarly, for the these TORC
measurements, we estimated (using the bootstrap method) that the SNR
of the {\em rank}-$1$ approximation is on average $3.4\pm 0.6$ times
higher than that of the raw measurement, while for the {\em rank}-$2$
and quadrant-separable approximations, the average gain in $SNR$ is
reduced to $2.0\pm 0.6$ and $1.9\pm 0.6$, respectively. Note that
these values are comparable to the $SNR_{cor}$ gain values $g$
employed in Section \ref{sec:ccs}. Although the {\em rank}-$1$
approximations have higher SNRs, which means that they remove
proportionally more noise than signal from the measurements, the
proportion of signal removed (as gauged by $\beta_{SVD}$) is
unacceptably high for many STRFs.

In order to cross-check the results obtained from $\beta_{SVD}$, we
recomputed $\alpha_{SVD}$ from the SVD approximations (denoted by
$\hat{\alpha}_{SVD}$), rather than from the raw measurements. For
example, if the quadrant-separable approximation is indeed a complete
and relatively error-free version of the true STRF, then computing
$\hat{\alpha}_{SVD}^{(1)}$ and $\hat{\alpha}_{SVD}^{(2)}$ from it
should yield results close to the corresponding $\beta_{SVD}^{(1)}$
and $\beta_{SVD}^{(2)}$ (from the raw STRF measurement). Similarly,
computing $\hat{\alpha}_{SVD}^{(QS)}$ from the the {\em rank}-$2$
approximation should yield a result close to $\beta_{SVD}^{(QS)}$.
These $\hat{\alpha}_{SVD}$'s are plotted in Figures \ref{fig:svd3}J
through L versus $SNR_{cor}$ for both the TORC and STWN measurements.
With respect to the original $\alpha_{SVD}$'s in \ref{fig:svd3}A
through C, they are more tolerant to changes in $SNR_{cor}$ over a
wider range of $SNR_{cor}$'s. When $SNR_{cor}$ is above $2$, these
$\hat{\alpha}_{SVD}$'s are indeed closely matched to the corresponding
$\beta_{SVD}$'s, as Figures \ref{fig:svd3}M through O attest. When
$SNR_{cor}$ drops below $1$, the $\hat{\alpha}_{SVD}$'s rapidly
increase and lose their correspondence with $\beta_{SVD}$, presumably
because the assumption that the SVD approximations are complete and
error-free rapidly breaks down.

In this section, we have concentrated on the TORC measurements. They
are ideal in that they produced low levels of systematic error and a
wide range of $SNR_{cor}$'s. The STWN measurements were less than
ideal in that $SNR_{cor}$ was limited below $2$. In Section
\ref{sec:syserr}, this was found to be chiefly due to high levels of
stimulus-induced systematic error; indeed $\beta_{SVD}$ was grossly
inflated in these measurements, rendering it no more illuminating than
$\alpha_{SVD}$ (not shown). Nevertheless, over the range of
$SNR_{cor}$ that they can be compared, the distributions of
$\alpha_{SVD}$ in Figures \ref{fig:svd3}A through C and
$\hat{\alpha}_{SVD}$ in J through L were very similar for the STWN and
TORC measurements. Moreover, from Section \ref{sec:ccs}, the SVD
approximations of STWN and TORC measurements were increasingly well
matched as the error level dropped. Therefore, the available evidence
supports the hypothesis that, for a given level of measurement error,
the STWN results and TORC results are equivalent, but the STWN results
are much more error prone. The dynamic-ripple results were less than
ideal in that STRF$_{{DR}}$ is quadrant-separable by
construction. Additionally, it contains non-uniformly distributed
errors \citep{dep1}, which complicates both the SVD \citep{stewart1} and
the interpretation of $SNR_{cor}$. With this caveat, we note that the
distribution (although not the range) of $\beta_{SVD}^{(1)}$ was
skewed toward somewhat higher values in the dynamic-ripple
measurements. For instance, $\beta_{SVD}^{(1)}$ exceeded $10\%$ in
$61\%$ of STRF$_{{DR}}$'s versus $45\%$ of
STRF$_{{TORC}}$'s. Still, $\beta_{SVD}^{(2)}$ was below $5\%$ in
$91\%$ of STRF$_{{DR}}$'s; the indications were that most
STRF$_{{DR}}$'s were still well described by {\em rank}-$2$
approximations.

In summary, the optimal SVD approximation of an STRF measurement with
a sufficiently low error level (e.g., $SNR_{cor}$ above $2$) does well
describe the STRF, in that it preserves {\em at least} $90\%$ of the
STRF's power. Therefore, we can be confident that if the SVD
approximations of two STRF measurements are well matched, so are the
corresponding STRFs. However, when there exist higher levels of
measurement error, this is no longer guaranteed to be the case,
particularly for STRFs that contain a significant non-separable
component. Overall, around $60\%$ of the TORC measurements were well
described as being separable. The rest were better served by both {\em
rank}-$2$ and quadrant-separable approximations, which were
essentially identical. To the extent that they could be compared, the
STWN and dynamic-ripple measurements produced similar results.

\section{Discussion}
\label{sec:discuss}

A pseudo-random exploration of
the space of spectrotemporal patterns, fostered by the traditional methodology of 
reverse
correlation, has been the basis of most previous STRF measurements.
Instead, we applied a deterministic and analytical reformulation of
reverse correlation, which is based upon the Fourier-series
description of dynamic spectra. One advantage of this approach
concerns experimental optimization: It enables us to restrict the
stimulus space to a minimal, discrete set of spectrotemporal patterns
(the spectrotemporal modulation frequencies, presented simultaneously
or individually). It also facilitates our understanding of measurement
errors and their various stimulus- and response-induced components. In
sum, it enables us to design stimuli that are efficient and
effective, while taking into account general knowledge of the STRF
structure, response nonlinearity and variability, and specific
laboratory constraints. A second advantage concerns experimental
evaluation: Since any given dynamic spectrum can be described by its
Fourier series, we can understand and quantify the performance of
different stimulation methods, even if they were devised within
different frameworks. Both of these advantages have been demonstrated
in this study, where we have measured STRFs of AI neurons with three
very different types of stimuli.

We now discuss the major empirical results of this study.

\subsection{Linearity}

The most striking finding is that when the STRF of an AI neuron is
successfully measured with different types of stimuli, the results are very
similar. The STRFs themselves exhibit a high degree of richness and
diversity across neurons. The three types of stimuli used, Dynamic Ripples,
TORCs, and STWN differ greatly in their spectrotemporal characteristics and
statistics (c.f. Figures \ref{fig:as}, and top panels in
\ref{fig:dr}A,B,C), and indeed they sound quite distinct from one another.
Great differences even exist between stimuli of a given type (except for
STWNs, which all sound noise-like). That STRFs measured from such widely
different stimuli are so similar speaks to the significance and robustness
of the linearity of neurons' responses with respect to the dynamic spectra
of stimuli.  Strong nonlinear system behavior would almost surely interfere
with the STRF measurements, not allowing the STRFs generated from such
different stimuli to have such large correlation coefficients (except
trivial cases such as static nonlinearities, e.g., rectification). The
correlation coefficients are especially large considering that the STRF
measurements contain large low-power regions (error-prone even after the
SVD), and furthermore compared measurements were often made over an hour
apart.

\subsection{Efficacy of the stimuli}

Although, when successful, they lead to very similar STRF
measurements, the three types of stimuli differ in their rates of
success. Success is achieved when the STRF measurement contains
sufficiently low levels of both non-systematic and systematic errors,
reflected by the measures of $SNR$ (using only non-systematic error)
and $SNR_{cor}$ (including systematic error). Non-systematic errors,
caused by response variability, are reduced when the modulations in
the stimulus are more powerful (evoking stronger modulations in the
response relative to the average spike rate), and also by averaging
the results from stimuli with identical spectrotemporal statistics.
Systematic errors, caused when multiple stimulus components evoke
interfering response components (either linearly or nonlinearly), are
reduced by careful stimulus design, or by averaging the results from
stimuli with identical spectrotemporal statistics (but different
individual characteristics). Note that all of the stimulus types used
had approximately the same total presentation duration.

On balance, the stimuli that gave the best results were TORCs, which
benefitted from careful stimulus design and relatively strong responses,
due to the restricted number of spectrotemporal modulation frequencies in
each stimulus.  As a result, we have noted that usable STRF measurements
could have been obtained after presenting one sweep of each TORC stimulus
(taking about $3$ minutes), a fact that we intend to exploit in the future.
STWN, while strongly motivated by the traditional reverse correlation
methodology, gave STRFs with substantially more systematic error than
TORCs. While both stimuli are capable of giving STRFs with high $SNR$, the
STWN results in substantially poorer $SNR_{cor}$. This is most cleanly seen
by comparing figure panels \ref{fig:snreff}A and \ref{fig:snreff}D: both
stimulus types give STRFs with $SNR$ as high as $30$, but STWN generated
STRFs have $SNR_{cor}$ that saturate below $2$, while TORC generated STRFs
have $SNR_{cor}$ saturating at substantially higher values.

Although the dynamic-ripple stimuli produce the most reliable results
(highest $SNR$), they suffered a fundamental flaw: Too many stimuli
were required to measure the full MTF$_{ST}$ (and hence its
STRF), and so measurements were restricted a subset of stimuli
required if the MTF$_{ST}$ is quadrant-separable. This is
problematic for two main reasons. First, it makes it impossible to
assess the quadrant-separability assumption directly. Although
quadrant-separability holds in (ketamine-anesthetized) AI, there may
be other neuronal populations or experimental conditions for which it
doesn't. Second, the full MTF$_{ST}$ measurement is a more
complex (nonlinear) function of the individual stimulus-response
relationships. This complicates the evaluation of measurement errors,
and thus blurs the distinction between neural functionality and
methodological artifact. Indeed, the dynamic-ripple results had a few
subtle idiosyncrasies, including more non-separable STRFs, and
$SNR_{cor}$'s poorly correlated with other assays of measurement
errors ($SNR$) and STRF structure ($\alpha_{SVD}$, $\beta_{SVD}$).
However, since the measurements are so reliable, it may be feasible to
sacrifice some $SNR$ by reducing the number stimulus repetitions in
order to present all stimuli required to directly measure the full
$H$ \citep{versnel02}.

Finally, we note that the TORC approach is not limited to the
particular stimuli used in this study. Any combination of
spectrotemporal modulation frequencies could exist in each stimulus,
provided that they are temporally orthogonal. Therefore one can
produce ``super'' TORCs, using fewer (but longer-duration) stimuli,
each of which contains many spectrotemporal modulation frequencies
\citep{robust}. These stimuli are more noise like, but benefit from a
lack of stimulus-induced systematic measurement errors in contrast to
the STWN stimuli. We are currently investigating the effectiveness of
such stimuli.

\subsection{The SVD: error reduction and signal loss}

In this paper, we used the SVD to reduce errors in the STRF
measurements. The SVD is ideally suited for use with the STRFs
measured here, because their SVD is strongly dominated by the lowest
order terms; that is, they are well approximated by a small number of
fully separable ({\em rank}-$1$) matrices. When such STRFs are
perturbed by unstructured errors, the SVD is still strongly dominated
by the lowest order terms, and has a well-understood contribution from
higher order terms. The boundary between the low order (high signal,
low error) and high order (low signal, high error) terms is not known
{\em a priori}, but is well understood from signal detection theory.
The upshot is that truncating the SVD series of an STRF at low order
is an efficient and well-understood way of increasing $SNR$ while
minimizing loss of signal.

Of the STRF measurements that were suitably error-free, more than half
were not only optimally approximated but well approximated (as
reflected by $\beta_{SVD}$) by fully separable ({\em rank}-$1$)
matrices. These approximations reduced the error power by at least a
factor of $3$ while sacrificing less than a tenth of the signal power.
The rest of the STRFs required two SVD terms ({\em rank}-$2$
approximations); using only one SVD term ({\em rank}-$1$) would give
an incomplete view of the system functionality due to excessive signal
loss. The {\em rank}-$2$ approximations have somewhat diminished error
reduction, down to a factor of $2$. Alternatively, the quadrants of
the MTF$_{ST}$ could be approximated by fully separable matrices,
producing results very similar to the {\em rank}-$2$ approximations.
However, if the error level was too high (e.g., $SNR_{cor}$ below
$1$), the optimal SVD approximations no longer reliably achieved both
significant error reduction and adequate signal retention. The error
level should always be considered when interpreting the results of the
SVD.

\subsection{The SVD: functional implications}

It is intriguing that STRFs are equally well described by {\em
rank}-$2$ and quadrant-separable approximations (see Figures
\ref{fig:svd3}H and I). These properties, each special in their own
right, do not necessarily imply one another. It turns out that if an
STRF is both {\em rank}-$2$ and quadrant-separable, special phase
relationships must exist, in either the temporal or spectral
dimensions (or both), between the separable matrices of the SVD or
equivalently the quadrants of the MTF$_{ST}$. It has been demonstrated
\citep{tempsep} that AI STRFs possess this property in the
temporal dimension (but not necessarily the spectral). This itself has
strong theoretical implications for the network connectivity of those
neurons.

\subsection{The error measures}

We found that incorporating systematic errors (otherwise known as {\em
bias}) into our consideration of the total measurement error level is
absolutely crucial for aligning the results from different types of
stimuli, and thus understanding the structure of an STRF measurement
(and the resulting SVD approximations, correlation coefficients, etc.)
independent of stimulus type. We used and analyzed two different
measures of error: $SNR$ and $SNR_{cor}$. $SNR$ is the more classical
but more limited of the two; $SNR$ is the ratio of the measured STRF
power to the measured STRF variance (square of the standard error).
This definition of $SNR$ (and its associated measure of error) is not
able to incorporate systematic error, however. In contrast,
$SNR_{cor}$ does incorporate systematic error. $SNR_{cor}$ is the
ratio of measured STRF power to measured non-STRF power (e.g., the
power in the spectrotemporal region where the underlying STRF is
expected to have near-zero power).

The only problem with $SNR_{cor}$ is that it requires assumptions
about the structure of the errors and the STRF, which may not apply to
all STRFs and stimuli. In particular, we assumed (based primarily on
observations) that errors are evenly distributed over the
measurements, and that the STRF power is near zero for $\tau$ above
$125$ ms (using negative $\tau$'s is no different since the stimuli
were periodic). The usefulness and predictability of $SNR_{cor}$
demonstrated that these assumptions largely held for the TORC and STWN
measurements. This was not the case for the dynamic-ripple
measurements, however, likely due to a combination of response
nonlinearity and the nonlinearity of the STRF measurement itself,
which distributes the errors non-uniformly in the spectrotemporal (and
modulation frequency) domain. It will be even more useful in the
future to devise measures of the systematic errors that are less
dependent on the structure of the STRF measurement.

\subsection{Response variability}

In our investigation of non-systematic errors in the STRF
measurements, several observations concerning the variability of AI
responses have interesting functional implications. For example, the
fact that the response variance could be linearly predicted from the average
spike rate in a nearly stimulus-independent manner points to a
Poisson-like spike generation mechanism, which has been vigorously
investigated in the visual system \citep{shadlen}. Additionally, we
found (see Figure \ref{fig:snr}C) that while neurons with higher-power
STRFs (higher $P$) tended to fire more spikes (higher $\bar{r}$, as
might be expected from a linear-plus-rectification response model), a
range of average spike rates were still observed for any given STRF
power. Neurons with the lowest spike rates (for a given $P$)
corresponded to the highest-SNR STRFs, and had the sharpest, most
phase-locked responses of the population (not shown). Neurons with the
highest spike rates often had seemingly random responses and
poor-quality STRF measurements. We will consider the origins and
implications of such behavior more carefully in future studies.

\subsection{Related studies}

Other recent studies have also addressed the similarity of STRF
measurements with different types of stimuli, albeit in different
auditory loci. Escab\'{\i} and Schreiner \citep{escabi1} measured
STRFs in cat inferior colliculus (IC) with stochastic stimuli that in
some respects resemble the dynamic-ripple stimuli and STWN used here.
While their results largely agree with ours, they singled out a small
group of neurons that either exhibited extremely selective and
phase-locked responses to the dynamic-ripple-like stimuli but were
unresponsive to the STNW-like stimuli (type-II neurons), or exhibited
non-phase-locked nonlinear responses to both stimuli (type-III
neurons). As discussed above, in AI we also find that neurons'
responses can be extremely sparse and yet yield significant STRFs
(like their type-II neurons). However, we did not observe two distinct
populations of neurons; rather, the degree of phase locking in
response to all stimuli ranged over a continuum. In addition, some AI
neurons exhibited significant spike rates but poor STRF measurements
(like their type-III neurons). Although we have not yet found a
nonlinear relationship between these responses and the dynamic spectra
of the stimuli, we can not yet rule out that possibility. In another
study, Theunissen et. al. \citep{theunissen} measured STRFs in the
zebra finch auditory forebrain in response to random tone sequences
and bird songs, and used the STRF from one stimulus to predict the
responses to the other. They found small but significant differences
in the cross-predictability of the responses, which was poor overall.
These differences either reflect differences in the STRF-measurement
method (which was implemented as a nonlinear function of the responses), or
more probably reflect a higher degree of nonlinearity in the responses
of neurons in the avian auditory forebrain with respect to mammalian
AI (but see \citep{schafer}, who reported a higher degree of linearity and predictability).

\subsection{Nonlinearity}

This article has been concerned with nonlinearities only insofar as
they interfere with the STRF measurement, and methods were invoked to
reduce this interference (e.g., the inverse-repeat method). Other
methods are also available, such as more carefully choosing the
temporal modulation frequencies in the TORCs, so that the nonlinear
distortion products are also orthogonal to the linear response (a la
\citep{victor}). That is not so say that nonlinearities form an
insignificant part of the AI response, merely that linearity is
important, strong, and robust to changing stimulus conditions, and
therefore forms an sturdy foundation upon which the study of auditory
cortical processing can be based, even in its nonlinear aspects.

We are currently investigating several anticipated nonlinearities.
These include the nonlinear transformation of responses occurs at the
thalamo-cortical depressing synapse, which contributes a rapid
adaptation of onset responses towards a steady state within a few tens
of milliseconds \citep{denham, kowalski1, phillips, heil} (we
considered only the steady-state response in this study).
Additionally, we have observed that when stimuli contain both low and
high modulation frequencies, AI responses can phase lock to much
higher frequencies than previously expected (e.g., $100-200$ Hz)
\citep{elhilali}. Similar effects have been observed in the visual
system \citep{bair, reid, chance1}. In our stimuli, these high
modulation frequencies result from interactions between unresolved AM
tones (that fall within the bandwidth of the same cochlear filter),
even though they were not part of the target dynamic spectrum (and
therefore did not contribute to the STRF measurement). A third
nonlinearity is the potential dependence of responses on the bandwidth
of the stimulus.  Broadband sustained stimuli (such as the ripples,
TORCs, and STWN) likely bias cortical cells in a manner different from
that of narrowband or transient stimuli such as tones and clicks.
Consequently, predicting details of tone and click responses from the
STRF may prove sometimes problematic \citep{kowalski2, theunissen}.
However, this nonlinearity is irrelevant when the focus is on
comparing STRFs derived from similarly broadband and sustained
stimuli, as is the case in this paper. Yet another important source of
nonlinear effects are static nonlinearities (e.g., rectification,
response saturation) with respect to stimulus level and contrast. By
fixing stimulus contrast at near maximum ($90\%$), and the absolute
level at an intermediate value (e.g., based on the rate-level function
\citep{kowalski2} we have managed to obtain reliable reproducible
results from a sizable proportion of cells in A1. Finally, there are
fundamental nonlinearities that we have not yet convincingly observed
in AI responses, such as units analogous to the complex cells of the
visual cortex \citep{devalois}. Nevertheless, it is likely that a
significant proportion of the very low SNR STRFs observed in this
study belong to cells that would be classified as nonlinear in that
they either phase-lock poorly to our stimuli or respond to more
complex patterns that we have not been able to probe (e.g., see
\citep{escabi1}).

\section*{Acknowledgments}

D.J.K. was supported at ETH Z\"{u}rich by the Caltech ERC-INI
collaboration grant 0-23819-01 from the Koerber Foundation. D.A.D. was
supported by the National Institute on Deafness and Other Communication
Disorders, National Institutes of Health, Grant R01 DC005937. S.A.S.
was supported by Office of Naval Research Multidisciplinary University
Research Initiative, Grant N00014-97-1-0501. 



\bibliography{abbrev,torcphys}

\end{spacing}

\end{document}